\documentclass[a4paper,12pt]{article}
\usepackage{amssymb,amsmath,mathrsfs}
\usepackage[usenames,dvipsnames]{xcolor}
\usepackage{hyperref, comment}
\usepackage{tensor}
\usepackage{graphicx}
\usepackage[left=1.2in,top=1in,right=1.2in,bottom=1in,headheight=0.8in,foot=0.5in]{geometry}
\usepackage[nodisplayskipstretch]{setspace} 
\usepackage[utf8]{inputenc}
\usepackage[greek,english]{babel}
\usepackage{lmodern}
\usepackage[width=\textwidth]{caption}
\usepackage{pifont}% http://ctan.org/pkg/pifont
\usepackage{xcolor}
\usepackage{amsmath,latexsym}
\usepackage{outlines}
\usepackage[textsize=tiny]{todonotes}

\usepackage{arabtex}
\usepackage{cjhebrew}
\usepackage[indentafter]{titlesec}
\titleformat{name=\section}{}{\thetitle.}{0.8em}{\centering\scshape}
\titleformat{name=\subsection}[runin]{}{\thetitle.}{0.5em}{\bfseries}[.]
\titleformat{name=\subsubsection}[runin]{}{\thetitle.}{0.5em}{\itshape}[.]
\titleformat{name=\paragraph,numberless}[runin]{}{}{0em}{}[.]
\titlespacing{\paragraph}{0em}{0em}{0.5em}
\titleformat{name=\subparagraph,numberless}[runin]{}{}{0em}{}[.]
\titlespacing{\subparagraph}{0em}{0em}{0.5em}

\usepackage[style=ext-authoryear-comp,maxcitenames=2,uniquename=false,backend=biber,uniquelist=false,articlein=false]{biblatex}
\DeclareNameAlias{sortname}{last-first}

\newcommand{\nocontentsline}[3]{}
\newcommand{\tocless}[2]{\bgroup\let\addcontentsline=\nocontentsline#1{#2}\egroup}
\hypersetup{
	colorlinks=true,         
	linkcolor=Black,          
	citecolor=MidnightBlue,
	urlcolor=MidnightBlue            
 }

\title{Methodological Reflections on the MOND/Dark Matter Debate}
\date{}

\begin{document}

\maketitle
\vspace{-.5in}

\begin{center}
\author{Patrick M. Duerr\footnote{Martin Buber Society of Fellows for Research in the Humanities and Social Sciences, Hebrew University of Jerusalem, IL. Email:\  patrick-duerr@gmx.de.}\footnote{Faculty of Philosophy, University of Oxford, UK. Email: patrick.duerr@philosophy.ox.ac.uk.} and William J.~Wolf\footnote{Faculty of Philosophy, University of Oxford, UK. Email:\ william.wolf@philosophy.ox.ac.uk, williamjwolf2@gmail.com.}}
\end{center}\vspace{.1in}

\begin{center}
Forthcoming in: \textit{Studies in History and Philosophy of Science}
\end{center}\vspace{.1in}

\begin{abstract}
\noindent The paper re-examines the principal methodological questions, arising in the debate over the cosmological standard model's postulate of Dark Matter vs.\ rivalling proposals that modify standard (Newtonian and general-relativistic) gravitational theory, the so-called Modified Newtonian Dynamics (MOND) and its subsequent extensions. What to make of such seemingly radical challenges of cosmological orthodoxy?
In the first part of our paper, we assess MONDian theories through the lens of key ideas of major 20th century philosophers of science (Popper, Kuhn, Lakatos, and Laudan), thereby rectifying widespread misconceptions and misapplications of these ideas common in the pertinent MOND-related literature. None of these classical methodological frameworks, which render precise and systematise the more intuitive judgements prevalent in the scientific community, yields a favourable verdict on MOND and its successors--contrary to claims in the MOND-related literature by some of these theories' advocates; the respective theory appraisals are largely damning. Drawing on these insights, the paper's second part zooms in on the most common complaint about MONDian theories, their ad-hocness. We demonstrate how the recent coherentist model of ad-hocness captures, and fleshes out, the underlying---but too often insufficiently articulated---hunches underlying this critique. MONDian theories indeed come out as severely ad hoc: they do not cohere well with either theoretical or empirical-factual background knowledge. In fact, as our complementary comparison with the cosmological standard model's Dark Matter postulate shows, with respect to ad-hocness, MONDian theories fare worse than the cosmological standard model. 
\end{abstract}
\newpage

%\tableofcontents

\begingroup
\makeatletter
\parskip\z@skip
\tableofcontents
\endgroup

\setstretch{1.2}

\section{Introduction}\label{intro}

\noindent According to the standard (``$\Lambda$CDM") model of cosmology (see e.g.\ \textcite[Ch.1]{Dodelson} for a didactic review), we don’t know what ca.\ 85\% of matter in the cosmos is made up of. The lion’s share of matter in our universe seems to belong to Dark Matter, an enigmatic and novel kind of matter; unlike ordinary (``baryonic”) matter, it’s not composed of quarks. Otherwise, the term serves as little more than a fig leaf for ``our ignorance regarding most of the matter in the universe” (\textcite{AviSciAmerican}; see also \textcite{Martens2021-MARDMR-5} for an illuminating analysis). To-date, attempts to detect Dark Matter via means \textit{other} than its gravitational effects (so-called direct detection experiments in particular) have failed, meaning that we know almost nothing of its other properties. Yet, most physicists firmly believe in the existence of Dark Matter on the basis of multiple forms of evidence, from---inter alia---motions of galaxies, gravitational lensing, or high-precision data of the Cosmic Microwave Background.

Advocates of MOND, an alternative theory of gravity, aspire to draw a less dark picture of the world by dispensing with Dark Matter.\footnote{There also exist hybrid proposals, such as that in \textcite{Berezhiani:2015pia}, which combine MOND-like modifications of gravity and Dark Matter (as well as those that blur the distinction between the two). We'll set aside these ``ugly solutions" \parencite{Vanderburgh2005-VANTMV-2,Vanderburgh2014-VANOTI-4}: although nothing conceptually or empirically precludes their in-principle viability---and potential prospects for ultimate success!---they inherit the worst of both worlds with respect to this paper's methodological questions. Our focus will therefore be on the ``pure" cases (see however e.g.\ \textcite{Martens2020-MARDM-13}).} Various astrophysical phenomena that the cosmological standard model associates with Dark Matter, MOND attributes to a deviation from standard gravitational theory: postulating Dark Matter, on MOND, is thus an \textit{artefact} of using the wrong theory.

MOND indeed triumphs on galactic scales (see e.g.\ \textcite{McGaugh:2020ppt} for a survey). Here, it’s scarcely hyperbolic to extol MOND as even predictively superior to the cosmological standard model, with its Dark Matter postulate. Also beyond that regime, relativistic field-theoretic extensions of MOND have been mooted. They allow a comparison with the standard gravitational theory, General Relatity (GR), especially on cosmological scales. Here too, MONDian research has accomplished some successes. The results are more mixed, though, as we’ll discuss throughout this paper; several proposals have in fact been ruled out.

In light of these achievements---together with the disappointment over the non-detection of Dark Matter, as well as over ``small-scale controversies", i.e. anomalies at the level of galaxies (precisely in the regime where MOND  celebrates its successes)---one may well wonder how seriously to take MOND: how significantly does it challenge established gravitational physics? How does MOND fare vis-à-vis the standard model of cosmology? How does MOND comport with customary methodological criteria for theory appraisal and selection? Little surprise then that the debate over the status of MOND vs.\ the Dark Matter postulate has an unmistakably philosophical calibre. Several physicists who participate in it (on both sides) have proffered analyses and arguments of overtly philosophical hue. Of late, also a handful of philosophers of science have joined the fray.

The present paper will revisit some of the salient broader methodological concerns that MONDian research has prompted. They deserve greater and more explicit attention than they have so-far received. We'll fill some of the gaps we perceive in the extant literature. Following a brief review of MOND (§\ref{review}), the paper’s first part (§\ref{method}) re-assesses key methodological arguments that one regularly encounters. Critical comments are due on the frequent invocations of some classical authors, such as Popper or Kuhn. They are often misrepresented, occasionally even grossly distorted. Contrary to widespread claims in the literature, invoking those authors largely speaks \textit{against} MONDian research. The astute analyses of these towering figures---whatever the ultimate merits of their views---can be profitably used as magnifying glasses for the pertinent questions that a judicious assessment should heed.  

The paper's second part (§\ref{adhoc}) will build on the insights gained by those classical authors' perspectives. Adopting a \textit{contemporary} perspective, we'll scrutinise the most common aspersion on MOND: that it’s ad-hoc. Regrettably, crucial details are usually elided: how to cash out the ad-hocness, putatively instantiated by MOND? How does it compare to the ad-hocness of the Dark Matter postulate? More generally, what makes ad-hocness, MOND's in particular, methodologically rebarbative, as the charge insinuates? Why ought we to avoid it? We’ll show how a recent model of ad-hocness due to \textcite{Schindler2018-SCHACC-4}, based on relations of coherence, offers illuminating answers. Its application to MOND captures and systematises many of the sceptical sentiments about MOND (and its extensions) in the physics mainstream. Finally, we'll compare the situation of MOND regarding ad-hocness with that of the cosmological standard model’s Dark Matter postulate. Here, Schindler’s model likewise allows a nuanced methodological evaluation. For the most part, it again buttresses the prevailing verdict: in terms of ad-hocness, the cosmological standard model is to be preferred over MONDian alternatives.  

\section{A Brief Review of MOND}\label{review}

\noindent To set the stage for the upcoming analyses, let's recapitulate here the basics of Milgrom's original theory, MOND \parencite{1983ApJ...270..365M}. A discussion of its relativistic extensions will be postponed to §\ref{adhoc}.

MOND modifies Newtonian Gravity in the  regime of low accelerations. It involves two principal innovations vis-à-vis Newtonian Gravity:
\begin{enumerate}
\item [MOND$_1$.]
The modification of the Newtonian force equation\footnote{MOND can also be interpreted as a ``modified \textit{inertia} theory", see \textcite{Milgrom:2005mc} for details. But it has received only marginal attention in the physics literature (see e.g.\ \parencite[pp.52]{Merritt2020}). We'll therefore set it aside, and instead focus on MOND's standard, force-theoretic variant.} for gravity for a test body of (inertial) mass $m$:
\begin{equation}
    F_G =m\mu(x)a_{g,N}.
\end{equation}
Here, $F_G$ denotes the modified gravitational force (to be plugged into Newton's Second Law, $m a =F_G$), $a_{g,N}$ the standard Newtonian gravitational acceleration, and $\mu(x)$ a scalar function (constrained primarily by its asymptotic behaviour, see below) of the ratio $x := a_{g,N}/a_0$ between the Newtonian-gravitational acceleration and the constant $a_0$.  
\item [MOND$_2$.]
The introduction of a new constant of nature $a_0$ (with the dimensions of an acceleration). It delimits the Newtonian regime from the low-acceleration (``deep-MOND") regime, where MOND posits departures from Newtonian theory. The empirically preferred value for  $a_0$  is $a_0 \approx 1.2 \times 10^{-8}$  cm s$^{-2}$.
Only for low accelerations, $x \ll 1$ does MOND postulate deviations from Newtonian Gravity. The transition is smoothly interpolated via $\mu (x)$, a free function, chosen such that for $x \gg 1$, $\mu (x) \rightarrow 1$. 

\end{enumerate}

 MOND scores significant empirical successes in the intended regime (see e.g.\\ \textcite{Sanders:2002pf, Famaey:2011kh, Merritt2020}): 
 
 \begin{itemize}
     \item MOND is constructed so as to accommodate the asymptotically flat rotation curves of spiral galaxies. There exists a discrepancy between the visible mass in galaxies and galaxy clusters, and the mass inferred from their dynamical effects. Their observed rotational velocities flatten out more than one would expect, given standard Newtonian Gravity and the observed luminous matter. The on-set of this discrepancy occurs, universally, above the deep-MOND regime (i.e. for accelerations below $a_0$). To account for this discrepancy, standard Newtonian theory must  postulate halos of non-luminous (``Dark") matter beyond the visible matter distribution.
    \item MOND predicts a \textit{specific} form of the observed rotation curves of isolated galaxy (or galaxy cluster) masses (the so-called ``baryonic Tully-Fisher relation"): mass (or luminosity) and asymptotic rotational velocity are related via a universal power law. Such a law is prima facie surprising due to the stochastic elements involved in the formation and evolution of galaxies (including the Dark Matter halos that the $\Lambda$CDM postulates).
    \item For any point in a disk galaxy, MOND gives a unique relation between the observed acceleration and the Newtonian gravitational acceleration, as calculated from the galaxy's luminous mass. This has a particularly startling implication (``Renzo's rule", see e.g.\ \textcite{McGaugh:2014xfa} for details): to each feature in the galaxy's luminosity profile corresponds a feature in the rotation curve, and vice versa. Such correlation appears unlikely if rotation curves are produced by \textit{non-luminous} (i.e.\ Dark) matter---especially in galaxies whose mass budget we have reason to believe to be dominated by Dark Matter.
    \item The central surface brightness of galaxies is correlated with their mass surface density, as determined via the galaxies' dynamics. Once more, such a correlation is surprising especially in galaxies with low surface brightness that are supposed to be dominated by Dark Matter (i.e. non-luminous matter inferred from the galaxy's dynamics).  
 \end{itemize}
  In light of these successes, one may ponder how seriously to take MOND and its more recent extensions. Some authors, such as \textcite{Merritt2020, pittphilsci19913}, have proclaimed that those MONDian theories even \textit{surpass} the cosmological standard model: we are summoned to relinquish the latter in favour of the former! Is that a fair assessment? In what follows, we'll push back against this. Contrary to claims in the literature, we'll show that the criteria of classical methodological authors (Popper, Kuhn, Lakatos and Laudan) yield a bleak appraisal of MONDian theories (§\ref{method}). A contemporary perspective likewise spells doom: MONDian theories are significantly more ad-hoc, we submit, than the cosmological standard model, with its postulate of Dark Matter (§\ref{adhoc}).   
 
\section{Classical Methodology in the MOND/Dark Matter Debate}\label{method}

\noindent Classical philosophers of methodology are routinely cited in the MOND-related literature. Regrettably, misportrayals and misleading applications of their ideas abound. This section will rectify the most relevant ones. In this, we don't intend to indulge in matters of exegesis---let alone, peddle an apologia. Rather, the MOND/Dark Matter debate benefits from circumspect readings of those authors: their sophisticated reflections may profitably be harnessed as magnifying glasses for salient methodological issues at stake in the debate. We'll successively inspect pertinent ideas of Popper (§\ref{popper}), Kuhn (§\ref{kuhn}), Lakatos (§\ref{lakatos}), and Laudan (§\ref{laudan}). In our presentation we'll avail ourselves of a modicum of opportunism: from these authors' thinking we'll cull only those ideas that, to our minds, insightfully bear on the MOND debate. For more comprehensive discussions, and critical assessments, of their views we refer to the literature.

\subsection{Popper---Against \textit{Na\"{i}ve} Falsificationism}\label{popper}

What verdict does Popper's falsificationist methodology deliver on MOND? Despite contrary claims in the MOND-related literature, MONDian theories, on a Popperian analysis, are found to be falsified, and furthermore methodologically inferior.\footnote{As we'll see, RMOND, is the main relativistic extension of MOND that hasn't been falsified (see §\ref{rmond}). The Popperian verdict of methodological inferiority still applies.} Yet, the Popperian vantage point gives due credit to MONDian research: even if unsuccessful at offering convincing solutions, MONDian theories sharpen our understanding of intriguing \textit{problems} that MONDian research has foregrounded.    

Popper is arguably the most widely cited philosopher of science amongst scientists, and amongst astrophysicists and cosmologists in particular (see \textcite{doi:10.1177/0270467604270151, Kragh2012-KRATAE-42,doi:10.1177/002182861204300306, Kragh2013-KRATMP-2,2021kpsp.book...53K}; also in the MOND-related literature, he is frequently invoked (e.g.\ \textcite{Merritt:2017xeh, Merritt2020, pittphilsci19913} or \textcite{Lazutkina2021TheAO} for defences of MOND) along explicitly Popperian lines). Correcting misrepresentations of Popper's falsificationism, prevalent in this literature, goes beyond doing exegetical justice to ``the outstanding philosopher of the twentieth century" \parencite{Magee1973-MAGP-5}. Popper holds plenty of valuable insights also for the MOND/Dark Matter debate (pace \textcite{norton}); his views have a surprising---and oft-misprized---power. Their special appeal lies in their epistemological modesty and accordance with practice and intuitions in the scientific community (to be expanded on shortly). In several regards, practicing scientists are likely to hail Popper's falsificationism (as we, following \textcite{Andersson}, \textcite{Albert}, \textcite{Keuth}, and \textcite{Duerr2022} construe it) as orthodox methodology, reflected in scientific practice. 

Popper's falsificationism operates within a thorough-going fallibilism. Certainty in empirical science is utopian. Popper insists on the invariable tentativeness and revisability of all scientific inquiry; they call for perpetual critical scrutiny of new and old ideas. Popper forsakes all pretensions to \textit{conclusive} ascertainability of truth. This allows us to dispel a common objection to Popper (found e.g.\ in \textcite{2022FoPh...52...31A} or \textcite{Carroll2019-CARBFN}): since conclusive, strictly deductive falsification is unattainable (for Duhemian reasons, see below), the argument concludes, Popper's falsificationism is  practically useless at best, and vacuous at worst. 

Accusations of such ``na\"{i}ve falsificationism" \parencite{Lakatos1978-LAKTMO} miss the mark. Popper deems falsifications likewise invariably \textit{non}-conclusive (see e.g.\ \textcite[p.187]{Popper1983-POPRAT}; \textcite[p.19]{Popper2002}): he disavows strict (incontestable and certain) falsifications which pronounce a theory's death once and for all; crucial experiments in this strict sense don't exist. This follows from Popper's general fallibilism, stressing the conjectural nature of all knowledge, and his endorsement of holism in theory testing (``Duhem's Problem", see e.g.\ \textcite{Ivanova2021-IVADAH}, cf.\ \textcite{Ariew1984-ARITDT}). That is, when a theory is confronted with recalcitrant evidence, logic and fixed rules don't allow us to specify whether to blame the theory in question, or any of the background assumptions (which include those concerning the accuracy and interpretation of the prima facie disconfirming data). Consequently, falsifications are likewise only tentative and provisional \textit{hypotheses}---guesses about where in our ``net of hypotheses", our ``web of conjectures” \parencite[pp.XXXIII, our translation]{Popper2005}, to locate errors; definitive locations---that is, absolute refutations---are impossible. Popper stresses that what counts as a falsification isn't solely a matter of logic. Nor do logical relations between evidence and the theory suffice for falsificatory judgements.

At least the young Popper of \textit{Logik der Forschung} eschews all claims to truth for methodological questions. As far as evaluating scientific theories is concerned, Popper proposes, we can and should dispense with truth and claims to it: all we need (and may hope for) are pragmatic, per se truth-unrelated surrogate criteria for \textit{rationally preferring} certain scientific hypotheses over others.\footnote{This also mandates that we abstain from claims about the closeness to truth: we can neither attain truth, nor even quantify how probable our theories are; ignorance is our insuperable lot---to the extent that we can't even quantify our ignorance. Later, Popper introduced the idea of verisimilitude, supposed to provide a measure of approximation towards truth. Such a notion is at odds with the early Popper's thoroughgoing fallibilism; moreover, Popper's own technical elaborations of the idea---Popper's attempts to flesh out the idea beyond a vague intuition or metaphor---are known to have failed, see e.g.\ \textcite[Ch.5]{Keuth} for details. The main results of our analysis of MOND through the Popperian lens aren't affected by these changes in Popper's views.} 

Falsification, as Popper envisions it, has an essential communal dimension. What counts as a falsifier depends on the consensus of the scientific community\footnote{Far from an infallible source of knowledge, scientific judgements at a collective/group level are, for Popper, themselves subject to critical scrutiny. Popper's point is that, as \textit{a rule of thumb}, we have have pragmatic reasons to trust the collective expertise of the scientific community---or rather, \textit{generically}, scientists have nothing better to work with.} on the test statements---an agreement on which singular statements to regard as possible observational data that can, in principle, refute a given theory. Such a consensus (or rather: dominant view) at the group level arises from the acceptance of test statements at the individual level. For the latter, researchers exercise their discretion through their free scientific judgement. Neither logic nor evidence inexorably \textit{compel} it; scientific judgements lack apodeictic necessity. Yet, that scientists at bottom trust their instinct isn't supposed to invite arbitrariness. To select scientific theories based on their performance vis-à-vis those falsifiers, Popper argues, is rationally justifiable through pragmatic considerations---in full view of the absence of certainty about ultimate truths (see \textcite{Duerr2022}).

The most important methodological advice that Popper issues fortifies the scientific ethos:  the admonition to exercise the ``critical attitude" (see also \textcite{Duerr2022} for an interpretation of falsificationism as a virtue-based---as opposed to rule-based---methodology): ``to try as hard as we can to overthrow our solution, rather than defend it" \parencite[p.xix]{Popper2002}. \textcite[p.158]{Popper1994-POPTMO-3} explicitly and repeatedly ``sums up” his ``whole view of scientific method” as, ``very simply, that it systematizes the pre-scientific method of learning from our mistakes. It does so by the device called critical discussion.” Popper exhorts scientists to cultivate open-mindedness (and, conversely, abjure dogmatism) and relentless criticism in science. This is his antidote to subjectivism and relativism: it's incumbent on individual scientists to critically examine test statements before accepting them. 

On the one hand, this invests falsificationism with an attractive permissiveness and liberalism; it allows a flexibility, apposite vis-à-vis the complexity of real science. On the other hand, those seeking methodological guidance will hanker after more informative/specific advice regarding theory selection. On what criteria to evaluate scientific theories? 

Popper's answer---his primary, ``supreme [methodological] rule" \parencite[p.33]{Popper2002}: conceive bold theories, and keep those that withstand strict tests (op.cit., p.9).\footnote{This rule, capturing the methodological essence of falsificationism, may strike some as unoriginal, if not outright bland: to the ears of practicising scientists, at least today, it sounds like a commonplace. Popper acknowledges as much: ``the pronouncements of this theory are, as our examples show, for the most part conventions of a fairly obvious kind. Profound truths are not to be expected of methodology. Nevertheless it may help us in many cases to clarify the logical situation, and even to solve some far-reaching problems which have hitherto proved intractable." (ibid, p.33). Popper's goal in his falsificationist methodology is to render explicit and unpack the rough-and-ready methodological rules, constitutive of scientific practice---``the rules of the game of empirical science" (ibid, p.32).} That is  (cf.\ e.g.\ \textcite[Sect.5]{sep-popper}), first, we should compare ``the new theory with existing ones to determine whether it constitutes an advance upon them. If its explanatory success matches that of the existing theories, and it additionally explains some hitherto anomalous phenomenon or solves some hitherto unsolvable problems, it will be adopted as constituting an advance upon the existing theories." The preference for a theory's greater empirical content (i.e.\ the ``size"\footnote{For the sake of charity, the usage of ``size" here should be loose (but nonetheless intuitive). ``As Grünbaum and others have shown convincingly, the attempt to specify content measures for scientific theories is extremely problematic if not literally impossible" (\textcite[p.77]{Laudan1977-LAUPAI}, see also fn.16 therein); Popper's attempts likewise fall under this verdict. Given the messy nature of real science, we don't deem this vagueness particularly problematic.} of its consequences that admit of a confrontation with observations), and hence its greater predictive power, is supposed to vouchsafe the growth of our (conjectural) knowledge. 

A theory must also be put to empirical tests. This is the second step of Popper's methodological prescription (ibid.): we test a theory by confronting its conclusions with the basic statements, accepted by the scientific community. 
If its conclusions match the basic statements, the theory is corroborated. According to Popper, in line with his fallibilism, it's imperative not to conflate corroboration with verification (being \textit{proven} true).

What to do in the face of conflict between the basic statements and the theory's conclusions? How ought scientists to react to apparent counter-evidence? Popper \textit{prescinds from any definitive, universal answer}! Sometimes even the acceptance (or interpretation) of basic statements should be revised: observations might be statistical flukes, marred by confounding variables, or even fabricated; the measurement techniques might not be reliable in certain contexts, etc. The scientific community should then re-examine the assumptions that led to the acceptance of the pertinent basic statements; one must ascertain which of those assumptions to drop and how to correct it. 

As a defeasible, rough-and-ready \textit{rule of thumb}, Popper seems to counsel (cautious) conservativism: although not exempting any beliefs from critical analysis and potential revision as a matter of principle, he usually  recommends accepting the basic statements agreed on by the scientific community. 
Beyond that, he steers clear of further substantive advice; all material decisions are relegated\footnote{Of course, the scientific community is neither infallible nor an unambiguously delimited institution with a press office enunciating \textit{unanimous} viewpoints. 
Popper rests content with a rough-and-ready characterisation, commended by pragmatic considerations of expediency---in lieu of epistemological principles with more ambitious claims to truth.} to the scientists themselves. In particular, his methodology declines to prescribe whether one should pursue a more conservative strategy (i.e. to adjust the theory via relatively minor modifications) or a more revolutionary one (look for a bold, new theory) in response to the falsifying test (or ``basic") statements \parencite[Ch.7]{Andersson}; only fresh empirical tests can adjudicate the choice a posteriori.
For Popper, thus, “there is no method of discovery, which can advise us prior to empirical testing how to modify falsified or problematic theories” (op.cit, p.188; our translation).\footnote{The price of this flexibility and liberalism is, of course, limited informativeness. Popper offers meagre positive guidance for researchers in terms of specific methodological advice (rules). He leaves all substantive decisions of theory choice and construction up to the scientists themselves.
\textcite[p.102, our translation]{Anderl}; (see also \textcite[p.90]{Worrall1995-WORRIP-2}) notes a grave consequence: ``if, for every unsolved problem, every putative falsification, one were to jettison the current research agenda in toto, with all its assumptions, scientific progress would scarcely be conceivable. One would permanently have to start from scratch and would prematurely also those proposals that, with a little more effort, might perhaps have been successful after all." This defect philosophers of science tried to remedy; some of them will preoccupy us in the subsequent subsections.} 
Each bold new conjecture Popper enjoins us to assess afresh and in its own right---against the basic statements of the day (provided, of course, they also stand up to critical scrutiny). 
Popper merely admonishes scientists to remain critical and not to lapse into dogmatism. He urges the scientific community to incessantly challenge old ideas with severe tests, and to propose new and audacious theories, and to subject them to subsequent critical scrutiny. Otherwise, scientists should follow their noses.

We have now assembled the key ingredients of Popperian falsificationism. What do they signify for MONDian theories? We begin with three negative outcomes.
\begin{itemize}
\item \textit{On Popperian standards, both MOND and its most popular relativistic extension, TeVeS, come out as \textit{falsified}.}

There exists large consensus in the scientific community on the evidence for GR across many scales. This straddles data from the post-Newtonian regime (such as the solar system), the strong-gravity vicinity of neutron stars or black holes, and cosmological scales. In Popperian terminology, such data constitute basic statements. Relative to them, MOND is as dead as a dodo: as a non-relativistic theory, whose scope is limited to the slow-acceleration regime, MOND scuppers already on manifestations of leading-order corrections to Newtonian Gravity (see e.g.\ \textcite{Will:2014kxa}): observations of light deflection, perihelion precession, gravitational redshift, and gravitational waves, etc.\footnote{This Popperian result of MOND's falsification captures what appears to be the main objection to MOND from \textcite{Sabineblog} in her review of Merritt's (2020) monograph (see \textcite{pittphilsci18810} for criticism in the same vein).}

GR's application to cosmology, the standard (``$\Lambda$CDM") model (see §\ref{lcdm}), admits of several, strict tests in the Popperian sense. ``A good example of such a [successful prediction] is the CMB power spectrum. That was  predicted long before it was observed, and no additional parameters are needed to explain it" \parencite[p.19]{pittphilsci18810}.\footnote{see e.g.\ \textcite{Lopez-Corredoira2017-LPETAP} or \textcite{Ferreira:2019xrr} for recent reviews of cosmological tests of GR.} The cosmological standard model passes those tests with flying colours. Needless to say, lacunae and open problems persist. (More on this shortly.) 

TeVeS too comes out as falsified: the currently widely accepted basic statements associated with the direct detection of gravitational waves refute it \parencite{2022FoPh...52...31A}. Conflict between the original assumptions and the data might still be averted through modifications in the background hypotheses. As Popper accentuates, such ``immunisation stratagems" are always possible; but methodologically, they are suspect (see below).

\item \textit{Spurious declarations of the cosmological standard model's falsification:}

\textcite{Merritt:2017xeh, Merritt2020, pittphilsci19913} or \textcite{Kroupa:2012qj} aver that the anomalous MONDian phenomena falsify the standard model; this is (at best) vastly exaggerated (see also \textcite[p.10]{pittphilsci18810} for a detailed rebuttal). For Popper, anomalies don't necessarily damn a theory: scientific theories are imperfect; none is free from anomalies and (vis-à-vis its precursors typically) \textit{new} problems. It's one thing to acknowledge open problems of the standard model that are awaiting a satisfactory resolution. To proclaim these open problems as in principle insurmountable on the standard model goes too far. Where MOND displays explanatory advantages over the standard model (see \textcite[Sect.3]{Massimi:2018hzy} and \textcite{MartensKing} for critical assessments), it's controversial whether those problems are insuperable obstacles for the latter: for instance, they might be an artefact of idealisations and approximations (in particular, of suppressed baryonic ``feedback"), or, as in the case of the lithium problem (plausibly viewed, for instance, as a problem of \textit{nuclear} physics) they might not even be a problem for the \textit{cosmological} standard model of at all (see \textcite{Chan2019-CHAACO-33} for details). The scientific consensus seems to be that it's, at best, too early to say what to make of those open problems (see \textcite{DeBaerdemaeker2020-DEBJSS} for further details); thus, they \textit{don't} qualify as Popperian basic statements that could---let alone actually do---falsify the standard model.

\item

\textit{Methodologically, a Popperian will be leery of relativistic extensions of MOND.}

She may impugn their independent testability.
Doubts are indeed in order as to their novel predictive content: relativistic extensions of MOND don't seem to admit of severe tests in the sense of (corroborated) predictions that are highly unlikely in light of prior knowledge. They are overtly \textit{constructed} so as to (i) fit the available empirical data,  (ii) have a Newtonian limit that, for low accelerations, reduces to MOND,\footnote{Consequently, it would be disingenuous to vaunt those theories' achievements in the MOND regime as novel.} and (iii) reproduce the astrophysical and cosmological phenomena that the cosmological standard model can account for.\footnote{For \textit{some} phenomena the cosmological standard model likewise assumes \textit{some} auxiliary hypotheses classified as conventionalist stratagems---as does arguably every theory \textit{in praxi}. Plausible cases in point include assumptions about the distribution of Dark Matter halos (cf.\, for instance \textcite{McGaugh:2020ppt}). But ``(t)he situation is extremely complicated" (op.cit., p.228): it involves many poorly understood astrophysical details, and faces non-trivial numerical/modelling-related challenges. It's therefore not surprising---nor, to our minds, irrational---that the scientific community, in the main, ascribes greater significance to tests in \textit{other} domains, less bedeviled by the convolutions, characteristic of ``gastrophysics" (i.e.\ the astrophysics of galactic processes, including galaxy formation). Indeed, in those other domains, as we'll argue in §\ref{lcdm}, various phenomena constitute independent lines of evidence in favour of Dark Matter; its existence and approximate contribution to the observable universe's overall mass are thus well-supported.}

It's not far-fetched to descry herein an attempt to rescue the original idea behind MOND from refutation. If so, relativistic extensions would qualify as what \textcite[p.62]{Popper2002} reprimands as ``conventionalist stratagems"---methodologically suspect attempts to evade empirical refutation (cf.\ \textcite{Merritt:2017xeh}, who charges the \textit{standard model} of such ploys). Conventionalist stratagems denote dodges for ``immunising a theory from criticism" \parencite[p.1983]{Popper1983-POPRAT}: by resorting to them one flouts the critical attitude. 

Another feature of relativistic extensions of MOND makes them methodologically suspect for Popper: they are dauntingly complicated.\footnote{We are aware of only one author disputing this: talking about RMOND (see §\ref{rmond}), \textcite{merrittc} makes the baffling claim that RMOND is ``an almost minimal modification to Einstein’s theory” and that he ``can hardly imagine that any truly successful theory could be much simpler than [RMOND]”.} Both the number of variables, as well as the mathematical-syntactical form of their basic equations far exceeds those of GR. For \textcite[Ch.7]{Popper2002}, there is a close (pragmatic) relation between simplicity and testability: the simpler and less involved a theory, the easier to test it. Hence, the lack of simplicity and parsimony aggravates the difficulties of testing those theories; their complexity renders them knottier to evaluate, and therefore methodologically undesirable.\footnote{One might counter that the GR, together with its Dark Matter postulate, is more complicated, but in the \textit{matter sector}: the physics of the prerequisite Dark Matter turn out to be more complex than that of MOND. To our minds, this is unconvincing for GR is a theory of \textit{gravity}; it remains largely silent on the matter sector. If one wants to assess GR's testability in terms of parsimony and simplicity, we should therefore look to gravitational physics.}  

\end{itemize}
  
Alongside these negative results, a Popperian analysis of MONDian research also has some positive ones (cf.\ \textcite[p.22]{pittphilsci18810}). MOND arguably shouldn't be understood as a theory proper (more on this in §\ref{kuhn}); via its role as a ``phenomenological formula", it's more naturally viewed as a compact codification of empirically robust phenomenology. The cosmological standard model, at present, doesn't offer a fully satisfactory account of this phenomenology. This generates an extremely interesting \textit{``problem situation"} that would delight \parencite[p.50]{Popper1983-POPRAT} (cf.\ \textcite{Popper1994-POPTMO-3, Conjectures}, passim):\footnote{\textcite{Peebles2014DM} would undoubtedly agree.}  ``(i)f the universe is made of cold dark matter, why does MOND get any predictions right" \parencite[p.220]{2021SHPSA..88..220M}? This problem-situation spurs researchers on to resolve it---to search for a satisfactory explanation of its phenomenology, as well as to test GR in this regime. Through MONDian theorising, irrespective of its eventual success, we gain greater acquaintance with the problem (see e.g.\ \textcite[p.97]{Popper1994-POPTMO-3})---its scope, ramifications and, last but not least, knowledge of unsuccessful attempts to tackle it.\footnote{By the same token, a Popperian would surely also underscore how MONDian research has benefited, and continues to benefit, \textit{philosophy of science}, as amply attested to by the recent surge in publications on the topic---including a the first ever Special Issue of \textit{Studies in History and Philosophy of Science} \parencite{Martens:2022dsg}. Of special interest, to our minds, is a more general issue that the debate over MOND has drawn attention to: the epistemological status and methodological challenges of heavy reliance on simulations in astrophysics and cosmology (see e.g.\ \textcite{Jacquart2020-JACOSA-2, Gueguen2020-GUEORI}).} 

Especially noteworthy in this regard is the regime in which these phenomena occur: for low accelerations GR has indeed not yet been tested thoroughly. Our confidence in GR in this regime stems from extrapolating its adequacy across a wide spectrum above this regime (as well as from the absence of theoretical reasons to expect deviations in the regime in question). MOND's predictive successes, Popper would urge, should motivate tests of GR in this regime, and reconnoitre it theoretically with audacious new theories (as MONDian researchers are indeed doing).\footnote{Feyerabend has drawn attention to a relevant function of rivalling theories, to enhance our understanding of theories and their problem contexts through theory pluralism---in a manner arguably compatible with Popper's views (\textcite{Bschir2015-BSCFAP, KBrad2021-KBRFTP}). From this perspective, the existence of (empirically successful) MONDian rival theories can act as a magnifying glass for shortcomings of the received view of the day (cf.\ \textcite{Merritt2021-MERFRA}). (To recognise the enhanced testability afforded by rivalling theories that show some explanatory advantages isn't the same as to call the original theory \textit{ipso facto} thereby ``effectively refuted" or ``falsified" (op.cit.); to do so, to our minds, is merely rhetorical razzle-dazzle.) The ``Planes of Satellite Galaxies Problem" (the apparent preferential alignment of the Andromeda Galaxy's and the Milky Way's satellite galaxies), ``Missing Satellites Problem" (the standard model's apparent overprediction of satellite galaxies of the Milky Way) and the ``'Too big too fail'-Problem" (the standard model's overprediction of \textit{luminous} satellite galaxies) are cases in point.

Also outside of its domain of success, a rival theory spawns cognitive disquiet---highly desirable for a Popperian: scientists become increasingly aware of the ruling theory's blemishes and defects. The standard model's lithium problem (i.e. the mis-tallying of the predicted and observed frequency of lithium), the discrepancies in the measured values of the Hubble constant, and oddities in the CMB on large cosmic scales (such as the Cold Spot) are cases in point. These issues accrue greater attention. They trigger an intensified process of critical scrutiny of the received wisdom, and exploration of novel ideas---all developments that a Popperian will welcome.}   
    
Popper's liberalism and permissiveness licence---and certainly don't forbid---friends of MOND to \textit{further explore and study} MONDian theories. Popper trusts that scientists' intuition is the best guide to the future development of science. Notwithstanding the privileged epistemic status of the scientific community (regarding test statements), heterodox ideas have their place within Popper's falsificationism.   
Should a bold and corroborated MONDian theory emerge, the scientific community should take it seriously. At present, though, this scenario remains a pious hope. 

\subsection{Kuhn---Against \textit{Facile} Invocations of Paradigms}\label{kuhn}

Here, we'll argue against recurrent suggestions that MOND counts as a Kuhnian paradigm, superior to the general-relativistic paradigm; in fact we'll argue against the suggestion that MOND counts as a paradigm \textit{at all}. Kuhn's reflections on theory virtues, however, shed light on the less than enthusiastic reception of MONDian research within the scientific community. 

Allusions to \textcite{Kuhn1962-KUHTSO-3} loom large in the MOND-related literature (see e.g.\ \textcite{Sanders:2013nma, McGaugh:2014nsa, Milgrom:2019cle}; including explicitly philosophical commentaries, see e.g.\ \textcite{Lahav:2014aea}; \textcite[Ch.4.1]{Anderl}). Especially common are references to paradigms, a notion integral to Kuhn's model of the history of science. With the latter supplying a blue-print for scientific theory dynamics, such references rhetorically do double duty (particularly clear in \textcite[p.4]{Milgrom:2012mi} and \textcite{Milgrom:2019cle}, passim): on the one hand, they are supposed to lend indirect (further) support for MONDian theorising; on the other hand, they seem to work as a defence mechanism against allegedly unjust criticism. The two-pronged argument comprises:
\begin{enumerate}
    \item [K1.] 
    In their scientific achievements, MOND and its successors are strikingly similar to historical, ground-breaking theories. This ought to encourage us to entertain those theories as serious, potentially revolutionary rivals to the GR-based standard model of cosmology.
   \item [K2.]
    Their revolutionary potential has been underappreciated. The reasons for this neglect parallel those impeding the acceptance of previous scientific breakthroughs (e.g.\ Copernican geocentrism or GR).   
\end{enumerate}
The argument invites several responses.
First, the essentially \textit{analogical} reasoning, inherent in (K1), is ipso facto tenuous: similarities between a novel theory, and a time-honoured, illustrious one imply nothing about the former's methodological status. Perhaps the argument should be read more charitably: Kuhn's model of theory dynamics---taken as a law of historiography\footnote{We'll set aside doubts that such ``laws of history" actually exist (cf.\ \textcite{Popper1957-POPTPO-17}).}---subsumes MONDian research and its development. 

Also this reading falls prey to a straightforward line of attack---the rejection of a presupposition of (K1): one may impugn the adequacy of Kuhn's framework as a (universal) historiographical model (see also \textcite{Kuhn1978-KUHTET-3}). Does it satisfactorily capture the historical development of a science? The present consensus no longer endorses the Kuhnian framework, at least not in toto (see e.g.\ \textcite{Toulmin1970-TOUDTD}; \textcite{Laudan1986-LAUSCP}, \textcite{Laudan1988}, and \textcite{Hepburn2015-HEPSM} for (critical) assessments).\footnote{A main complaint about Kuhn's model targets its emphasis on discontinuities during revolutionary periods, and its converse de-emphasis of revolutionary innovation during periods of normal science (see e.g.\ \textcite[p.49]{Bird2000-BIRTK-4})} 

Yet another objection blocks the above arguments. One may baulk at them as naturalistic fallacies: indeed, it's controversial whether Kuhn's historiographical model is able to offer any methodological advice beyond the analogical reasoning at all. \textcite[p.198]{Feyerabend1970-FEYCFT} poignantly articulates the concern:\footnote{It generalises in fact to all so-called historicist approaches to philosophy of science (see e.g.\ \textcite{sep-rationality-historicist}).} ``whenever I read Kuhn, I am troubled by the following question: are we here presented with \textit{methodological prescriptions} which tell the scientist how to proceed; or are we given a \textit{description}, void of any evaluative element, of those activities which are generally called `scientific'? Kuhn's writings, it seems to me, do not lead to a straightforward answer".

But suppose that one somehow bridges this is/ought-gap; suppose, that is, that we can render plausible the inference from a Kuhnian description to a normative methodological evaluation. Still, the above argument can be assailed---by denying (K1): one may reject the premise that MONDian theories resemble other revolutionary theories (in Kuhn's model). Do MOND and the theories it engendered really fit the mould of the Kuhnian model? 
In particular, does Kuhn's notion of a ``paradigm" apply? Let's zoom in on this latter question. Contra e.g.\ Merritt (2020, passim), we answer it in the negative.

Recall the two main senses of ``paradigm". Its primary one denotes an exemplary or model solution to a set of problems or puzzles, such as Aristotle's \textgreek{Φυσικὴ ἀκρόασις} or Newton's \textit{Principia} and \textit{Optics}: the paradigm as exemplar is a concrete, impressive achievement of science (for our purposes: a successfully applied theory); its way of tackling a problem inspires fellow researchers to emulate it, and to extend its key ideas to other areas and problems.\footnote{NB: Not every solution to an empirical problem---nor every predictively successful regularity or even theory---counts as a Kuhnian exemplar (see \textcite[p.68]{Bird2000-BIRTK-4} for details).} Derived from the exemplar is a paradigm's secondary sense: the paradigm as a disciplinary matrix. It denotes a framework, or a package of methods, techniques, scientific practices etc. modelled on, and to some extent contained, in the exemplar.
 
MOND is an ill-suited contender for a paradigm in either sense for three reasons. First, MOND itself, in its formulation of §\ref{review}, is rarely considered a theory proper\footnote{Penrose's comment is characteristic: ``(a)lthough [MOND] seems to fit the facts remarkably well, there is as yet no coherent theory of this which makes good overall theoretical sense" \parencite[p.781]{Penrose}.}---even by its advocates. For instance, \textcite[p.101, our emphasis]{sanders_2016} brands it an ``empirically based \textit{algorithm}", and \textcite[p.3]{Milgrom:2014usa} refers to it as ``a phenomenological scheme". Its domain of validity (more precisely: the domain where it deviates from gravitational standard theory) is limited to galactic phenomena in the low-acceleration regime; it can't account for \textit{any} of the well-established post-Newtonian phenomenology, such as Mercury's perihelion, or cosmology. This impotence bears special weight in light of the existence of a rival theory, successfully covering these phenomena (cf.\ \textcite[p.77]{Kuhn1962-KUHTSO-3})---the standard gravitational theory, GR. It would be exaggerated to deplore GR's less satisfactory performance on galactic scales as ``a \textit{pronounced} failure in the normal problem-solving activity" \parencite[p.75]{Kuhn1962-KUHTSO-3} that would enthrone MOND as a paradigm (more on this in §\ref{lcdm}).  
Furthermore (as we'll expound in detail in §\ref{mond}), MOND violates fundamental theoretical principles (such as energy-conservation or the Strong Equivalence Principle). Despite certain predictive successes in its domain, it would be preposterous to liken MOND's achievements and those of the \textit{Principia} or the \textit{Optics}. 

Similar reservations apply to MOND's relativistic extensions (such as TeVeS or RMOND, see §\ref{adhoc}). They were devised in a laborious process of piecemeal theory construction and improvement. It would belie the painstaking trials and tribulations of this process, were one to tout those theories as \textit{paragon} solutions to puzzles in galactic astrophysics. They were procured through ingenuous technical \textit{tinkering}. \textcite{Carrollblog} scorns TeVeS (representative of other relativistic extensions of MOND) as ``an ungodly concatenation of random fields interacting in highly-specific but seemingly arbitrary ways". 

A second reason militates against identifying MOND with a paradigm: MOND differs from, say, Newton's \textit{Principia} in terms of its fecundity. In contrast to MOND, paradigms ``provide models from which spring coherent traditions of scientific research" \parencite[p.10]{Kuhn1962-KUHTSO-3}. Newton's \textit{Principia} sparked off a research agenda: it provides a mathematical language, a conceptual framework for physical theorising, and a general methodology for conducting research. By no stretch of the imagination can the same be said about MOND: besides postulating a universal acceleration scale on which gravity is supposed to depend, it makes no suggestions for further theory construction; it lacks any heuristic force that may guide or inspire fellow researchers to emulate the MONDian solution to some gravitational puzzles.\footnote{ \textcite{Milgrom:2009ee} often adduces spacetime scale invariance as an important and fertile constraint that MOND underwrites. An anonymous referee has made the  suggestion that this sounds similar to Einstein's (see \textcite[Ch.8.4.1.]{RBrown2005-RBRPRS-2}) characterisation of Lorentz Covariance as the Big Principle that encapsulates the gist of Special Relativity. We reject this intriguing analogy for two reason. First, unlike Lorentz covariance, scale-invariance doesn't enjoy the status of a robustly corroborated symmetry principle. In particular, in contrast to the universality of Lorentz covariance (in the absence of gravity), not only is scale-invariance limited to gravity; it becomes furthermore  ``visible" only in the low-acceleration (deep-MOND) regime. \textcite[p.3]{Milgrom:2016scu} tellingly writes: ``(t)he inspiration for this requirement [scale-invariance] stems from the observation that the rotational speeds [...] of test particles in circular orbits around spiral galaxies become $r$-dependent at large radii, $r$[...]". Secondly, scale-invariance has so-far turned out to be heuristically barren. This likewise stands in marked contrast to Special Relativity's Big Principle: inter alia, the latter produced after all the \textit{whole} of relativistic quantum mechanics! As \textcite[p.1]{Skordis:2020eui} state: ``[...]that [MOND] is scale invariant [...] has not yet led to a definitive conclusion as to how this invariance could lead to a MOND gravitational theory”. In short, scale-invariance, to our minds, is merely of interest, insofar as it's a symmetry of MOND, manifesting itself only in the deep-MOND regime; it can't lay claim to a more fundamental symmetry principle, which one should impose in further theory construction.}

This lack of fertility is closely related to a third dissimilarity between Newtonian Gravity and MOND: the lack of conceptual autonomy (or, to hark back to Kuhn's above-cited passage, the \textit{coherence} of the MONDian research tradition).\footnote{For the sake of the argument, let's gloss over the fact that Newtonian Gravity---MOND's most plausible ``rival" paradigm---was proposed almost 300 years ago. The latter has in the meantime been superseded. To present MOND \textit{and GR} as competing paradigms would be absurd. (So it would be to present \textit{a relativistic version} of MOND, and GR as rival paradigms: at present there simply \textit{is no} serious contender for a relativistic version of MOND, see §\ref{adhoc}.) MOND meets neither of Kuhn's conditions for paradigm replacement in \textcite[]{Kuhn1962-KUHTSO-3}(cf.\ Kuhn's view on progress, e.g.\ \textcite[Ch.6]{Bird2000-BIRTK-4}): ``(f)irst, the new candidate [paradigm] must seem to resolve some outstanding and generally recognized problem that can be met in no other way. Second, the new paradigm must promise to preserve a relatively large part of the concrete problem-solving ability that has accrued to science through its predecessors." \parencite[p.169]{Kuhn1962-KUHTSO-3} We'll therefore adopt the counter-historical scenario where the choice is between MOND and Newtonian Gravity.} MONDian theorising didn't evolve organically from a handful of powerful ideas (cf.\ \textcite[Sect.3]{MartensKing}).\footnote{cf.\ Einstein's (complementary) physical and mathematical strategies during his quest for GR, see \textcite{Janssen2007, Pitts2016-PITEPS}.} For a paradigm (in either of its senses) in its own right, MOND seems too \textit{parasitic} on Newtonian Gravity: \textit{all three} of MOND's basic postulates (ibid., Ch.4; see \textcite{Bekenstein:1984tv}) make essential reference to Newtonian gravity! MOND's appellation signals this reliance on its established rivalling theory: MOND was introduced as a modification of \textit{ Newtonian} gravitational dynamics. Rather than a paradigm in its own right, MOND is better viewed as an idea within the \textit{Newtonian} paradigm (that is, within the Newtonian paradigm as a \textit{disciplinary matrix})!\footnote{By the same token, Hall's 1894 modification of Newton's law of gravity through a correction to the Newtonian gravitational potential, capable of accommodating Mercury's perihelion advance \parencite[p.312]{SmithG}, isn't a paradigm \textit{sui generis}. Rather it's an idea, a desperate reaction to an anomaly, within the Newtonian disciplinary matrix.} 

Again, something similar applies to relativistic extensions of MOND. Perusing the ideas and principles employed in the construction of viable MONDian theories, one gets the impression that they are largely \textit{borrowed} from other theories. The desiderata and guiding principles that \textcite{Bekenstein:2004ne} lists are either directly imported from GR (as in the case of the Covariance Principle and the weak equivalence principle), or generic demands on garden-variety field theories (as in the case of causality constraints). Note in particular that, say, TeVeS explicitly uses, and supplements, GR's Einstein-Hilbert action. To our minds, one oughtn't to regard TeVeS's principles as constitutive of a paradigm \textit{sui generis}. Instead wouldn't it seem fairer to classify TeVeS as modifications\footnote{According to Bird's reconstruction of Kuhn, not every modification of a paradigm necessarily inaugurates a new paradigm; a paradigm can change \parencite[p.42]{Bird2000-BIRTK-4}.} of GR's paradigm (similar to, say, $f(R)$ Gravity or Brans-Dicke theories)?\footnote{Indeed, this is what \textcite[p.300]{Sanders:2002pf} seem to suggest.}

Denying MOND (or its successors) the status of a paradigm in the Kuhnian sense isn't to gainsay MOND's explanatory and predictive successes (in the slow-acceleration regime for weak gravitational fields). Our point is simply that not every new (or even every radical) and empirically successful idea counts as a Kuhnian paradigm. Ironically, and contrary to the (as we argued: multiply flawed) argument with which this subsection set out, an idea of Kuhn's illuminates why MOND has attracted so \textit{few} adherents in the scientific community---notwithstanding those explanatory and predictive successes. 

First, one might wonder why MOND is being studied (by at least some researchers) at all: its non-relativistic domain, one might think, renders it  anachronistic in an era where general-relativistic physics is flourishing. Kuhn's two-phase model of scientific change---alternating between ``normal" and ``revolutionary" science---gives an answer: desperate times clamour for desperate measures. The cosmological standard model is undoubtedly in a crisis as Kuhn envisioned it: dark matter phenomenology, and cosmic expansion (both in the very early (inflationary) universe and in late times) constitute some of its foremost anomalies. 
This, according to Kuhn, characterises the predicament of ``normal science" in its late, \textit{critical} stages, prior to the emergence of a new paradigm. In crises, Kuhn's model predicts an exuberant proliferation of proposals. Some of those attempts to grapple with the anomalies stray far from conventional wisdom. We indeed observe such a proliferation: the literature on alternative theories of gravity attests to it (see e.g.\ \textcite{Clifton:2011jh}). The existence of a multitude of (relativistic) alternatives to GR goes a long way to explain why MOND has remained a minority view: MOND is \textit{one} proposal amongst many---most of them well-motivated both theoretically and empirically (see e.g.\ \textcite[Ch.1]{Faraoni:2010pgm}). 

To our minds, however, Kuhn's most penetrating insights for the MOND debate lie elsewhere: in his emphasis on so-called theory virtues (cf.\ also \textcite{MartensKing}).\footnote{This aspect of Kuhn's thinking needn't wed one to his historiographical model. In fact, theory virtues have enjoyed much attention in recent philosophy of science (see e.g.\ \textcite{Carrier2013-CARVAO-3, Douglas2013-DOUTVO-2, Keas2017-KEASTT, schindler_2018, Schindler2022-SCHTVD}), largely independently of Kuhn's two-phase model of science.} \textcite{Kuhn1977-KUHOVJ} draws attention to ``values" (or ``virtues") that theories may instantiate: these features, taken as hallmarks of good theories, play a crucial role in how scientists evaluate, and accordingly deal with, a theory. Kuhn lists a sample of five such values: accuracy (agreement with experiment), scope (covering as wide a range as possible), internal consistency and compatibility with other accepted theories, simplicity and parsimony, and fruitfulness (the ability ``to disclose new phenomena or previously unnoted relationships among those already known", op.cit, p.322). These values are, according to Kuhn, universally shared by the scientific community---\textit{across} different paradigms. All scientists employ them as determinant factors for rejecting/accepting a theory or a hypothesis.

Despite consensus on those values as criteria for theory choice, Kuhn underscores, their concrete application typically \textit{doesn't} yield the same results. Variety persists for two reasons. First, scientists may understand each virtue differently. For instance, simplicity notoriously admits of different interpretations (see e.g.\ \textcite{Fitzpatrick2013-FITSIT-2}; and \textcite{Vanderburgh2014-VANQPE} and \textcite{MartensKing} for a discussion in the context of Dark Matter)---say, syntactic-formally or in terms of qualitative or quantitative ontological parsimony. Secondly, even when agreeing on their individual meaning, scientists may assign the virtues differential weights: different scientists are likely to \textit{rank} the individual virtues differently.  

Kuhn's stance on theory values straightforwardly carries over to MOND's reception in the scientific community: it plausibly accounts for MOND's marginality. On its turf, MOND earns a high rating on the individual virtues of accuracy and what Kuhn calls ``fertility" (i.e. the capacity for novel predictions). But with respect to simplicity and scope it gets a low rating. If the choice \textit{were} between MOND and Newtonian Gravity, one might expect many researchers (or at least more than at present) to opt for the former: for reasonable weightings of theory virtues, their trade-off would plausibly tip the balance in favour of MOND. But that dilemma is science fiction: \textit{today's} choice is, at best, between MOND and GR.\footnote{Isn't this comparison between GR and MOND, which will unsurprisingly favour GR, tantamount a strawman? Shouldn't the comparison be between GR and some relativistic version of MOND? Indeed it should. Alas, at present no satisfactory relativistic MONDian theory---let alone one that MONDians could boast of as exemplary (see our upcoming discussion of the peculiarities of those relativistic theories in §\ref{teves} and §\ref{rmond})---exists for such a comparison. Recall that \textcite[p.357]{Kuhn1977-KUHOVJ} holds empirical accuracy to be ``ultimately most nearly decisive of all the criteria". Being empirically inadequate, TeVeS thus seems out of the game.} (Given the plurality of alternatives to GR, even that binary choice skews the state of affairs in gravitational physics.) Only a strongly biased premium on \textit{accuracy within MOND's domain} would underwrite a preference for MOND: by any reasonable standards, in terms of scope (with GR being extremely well-tested across a wide range of applications, see e.g.\ \textcite{Will:2014kxa, will_2018}), simplicity, compatibility with background knowledge (viz.\ \textit{all} of relativistic physics!), and fruitfulness (see e.g.\ \textcite{Weinberg1972-WEIGAC}), GR trumps MOND. We'll resume a similar-spirited analysis from a contemporary perspective in §\ref{adhoc}.   

\subsection{Lakatos---MONDian Physics: a Progressive Research Programme?}\label{lakatos}

\textcite{Merritt2020} (cf.\ also \textcite{Lahav:2014aea}) presents an in-depth evaluation of MONDian research (including TeVeS) in terms of Lakatos' methodology of scientific research programmes (henceforth abbreviated as ``MSRP")). According to Merritt, the standard cosmological model is methodologically morbid; MONDian research, by contrast, has developed in a methodologically salubrious fashion (is ``progressive"); hence we should eliminate the former in favour of the latter. We'll voice three grievances regarding Merritt's analysis.
His application of Lakatosian methodology is, we submit, flawed:
\begin{enumerate}
    \item [L1.]
    One may contest that the sequence of MONDian theories proposed over time constitutes a research programme as Lakatos envisages it.
    \item [L2.]
    Even if we regard it as such, its ``positive heuristic" is essentially impotent. 
    \item [L3.]
    The sequence of MONDian theories fails to be progressive. 
\end{enumerate}
Also on Lakatosian standards, the standard model of cosmology with its Dark Matter postulate comes out as preferable to MOND-inspired theories. Moreover, to invoke Lakatos, as Merritt does,  to defend of MONDian research is perplexing: Lakatos' methodology expressly refrains from any forward-looking methodological advice (especially regarding scientific promise), and postpones judgement of a theory's acceptability until the far-distant future.

With his MSRP, Lakatos attempts to improve on Popper's critical rationalism in light of Kuhn's challenges (see e.g.\ \textcite{Carrier2002-CARESP}): how to salvage the \textit{rationality} of theory change across the history of change? Of chief relevance for our purposes is Lakatos' attempt, within this agenda, to emend Popper's criteria for theory rejection (or rather ``non-/unacceptance", cf.\ \textcite[p.4]{Barseghyan2017-BARHCA-4}): MSRP tries to articulate criteria for ascertaining when it's rational to eliminate a theory in favour of another.  

For Popper, the unit of methodological evaluation is an individual theory. Lakatos, by contradistinction, hones in on (temporal) \textit{sequences} of theories, ``research programmes". They denote ``the
sum of the various stages through which a leading idea passes" \parencite[p.51]{Larvor1998-LARLAI-3}, a diachronic series of theories that share certain central assumptions and basic principles. Lakatos calls these assumptions the programme's ``hard core"; they constitute the programme's ``leading idea"---its essential features.

In addition to the hard core, each theory in the programme involves a number of auxiliary hypotheses. They compose what Lakatos labels the ``protective belt". Scientists are typically reluctant---and, according to Lakatos, rightly so---to abandon a research programme's core even when running into anomalies. They treat a research programme's hard core as inviolable. The auxiliary assumptions surrounding it fulfil two main functions. One is to confer observational significance: on its own, the core is (usually) devoid of empirical consequences; only in conjunction with further assumptions does it acquire observational content. A suitable set of auxiliary assumptions can, Lakatos presumes, always guarantee empirical adequacy. Hence those auxiliary assumptions' second function: they form the core's ``protective belt" that staves off empirical refutation in the face of empirical anomalies. 

Within MSRP, such anomalies are resolved via modifications in the protective belt; the programme's core is left intact. Lakatos dubs this injunction to preserve the core the programme's ``negative heuristic". It's supplemented by a constructive counterpart, ``a powerful problem-solving machinery, which, with the help of sophisticated mathematical techniques, digests anomalies and even turns them into positive evidence" \parencite[p.4]{Lakatos1978-LAKTMO}. The research programme's ``\textit{positive} heuristic consists of a partially articulated set of suggestions or hints on how to change, develop the `refutable variants' of the research programme, how to modify, sophisticate, the `refutable' belt" (op.cit, p.50). The positive heuristic delineates a \textit{programme} for future research: it contains a ``vision" of how to further develop theories. In particular, it delimits the kinds of revisions (viz.\ alterations in the auxiliary assumptions) one may undertake in order to deal with anomalies. \textcite[p.69]{Worrall1978} helpfully suggests ``a list of some of the things a positive heuristic may include. [...] 
The positive heuristic may include mathematics---for example, how theoretical assumptions should be formulated so that consequences may be drawn from them will be guided by the available mathematics; the heuristic may include hints on how to deal with refutations if they arise (e.g.\ `Add a new epicycle!'); and it may include directions to exploit analogies with previously worked out theories [...]".

For Lakatos, research programmes not only evolve in response to anomalies. Wholesome research programmes have an internal impetus, encoded in the positive heuristic: ``(m)ature science consists of research programmes in which not only novel facts but, in an important sense, also novel auxiliary theories, are anticipated;
mature science---unlike pedestrian trial-and-error---has `heuristic power'. Let us remember that in the positive heuristic of a powerful programme there is, right at the start, a general outline of how to build the protective belts: this heuristic power generates the autonomy of theoretical
science" (op.cit., p.88).

Consider now a research programme $\mathcal{P}$, constituted by the sequence $\mathcal{P}:=< T_1$, ...$T_i$,...,$T_N>$. Lakatos next introduces four criteria for  methodologically healthy changes \textit{within} a given research programme (see \textcite[p.61]{Carrier2002-CARESP}):

\begin{itemize}
    \item[] \textit{(C1)}: The transition from the $i$-th stage to its successor, $T_i \rightarrow T_{i+1}$, conforms to the research programme's positive and negative heuristic.
    \item[] \textit{(C2)}: The transition preserves $T_i$'s corroborated\footnote{See \textcite[p.34]{Rott1994-ROTZWV} for a discussion of some of Lakatos' ambiguities. We follow Rott's reading.} empirical content: $T_{i+1}$ reproduces $T_i$'s empirical successes.
    \item[] \textit{(C3)}: $T_{i+1}$ ``predicts some novel, hitherto unexpected fact" \textcite[p.33]{Lakatos1978-LAKTMO}. 
    
    Here, the novelty demanded of the predictions can be spelt out in different ways (cf.\ \textcite[Ch.2]{Merritt2020}): on a ``temporal" (Carrier) reading, the phenomena themselves have not been known beforehand; on a ``comparative" reading, the phenomena must be unlikely or even forbidden in light of a rivalling theory; on an ``interpretative" understanding of novelty, it suffices if a novel interpretation is given of known phenomena. (We needn't embroil ourselves in which interpretation of novelty is most appropriate.) 
    \item[] \textit{(C4)}: Finally, those novel empirical consequences must be corroborated: the predicted phenomena must actually have been observed. 
\end{itemize}
Lakatos calls a research programme ``\textit{theoretically} progressive", if each of its stages satisfies these criteria \textit{(C1)-(C3)} (i.e. if they hold for all $T_i$'s of the research programme). Note that such a healthy development requires that the programme evolve as autonomously as possible, fuelled by its positive-heuristic power: ``a healthy research programme is driven [...] principally by its heuristic. So long as there are some `dramatic' empirical results, and a steady supply of the kind of problem for which the techniques of the heuristic are effective, the programme can ignore anomalies. It is only when the heuristic runs out of steam that the anomalies have to be taken seriously" \parencite[p.55]{Larvor1998-LARLAI-3}.
For an \textit{empirically} progressive research programme, its theories must also satisfy \textit{(C4)}: the ``excess empirical content" of every $T_{i+1}$ over $T_i$'s empirical content must be corroborated, at least partially \parencite[p.110]{Lakatos1978-LAKTMO}. Finally, Lakatos labels a research programme ``progressive" simpliciter, iff it's both theoretically and empirically progressive; else he brands it ``degenerate". 

These \textit{intra}-programmatic standards of theory evaluation form MSRP's basis for adjudicating \textit{between} rivalling research programmes: progressive research programmes are methodologically preferable to degenerate ones, which lag behind in their theoretical and empirical accomplishments, ``running fast to catch up with their rivals" (op.cit, p.6). That is, for Lakatos, the transition from one research programme, $\mathcal{P}_1$, to another, $\mathcal{P}_2$---i.e.\ $\mathcal{P}_1$'s elimination  in favour of $\mathcal{P}_2$---is rational if $\mathcal{P}_1$ is degenerate, while $\mathcal{P}_2$ isn't. If both programmes are progressive, it seems plausible that, according to Lakatos, the extent to which \textit{(C2)} and \textit{(C4)} are satisfied---essentially, the ``size" of their corroborated empirical content--break the tie (but Lakatos doesn't elaborate on this situation, cf.\ \textcite[fn.6]{Carrier2002-CARESP}).     

Against this background of Lakatos' MSRP, we'll now argue that appealing to Lakatos backfires for advocates of MOND on multiple fronts:

\begin{itemize}
    
    \item Does MOND-related research fit the mould of Lakatosian research programmes? Reasons for doubt parallel those why, to our minds, the Kuhnian notion of a paradigm doesn't apply to MONDian research: it's difficult to fathom what MONDian theories have in common apart from very general desiderata and, of course, MONDian itself as a limit; it's obscure what might constitute the programme's ``hard core". The already-mentioned principles on Bekenstein's list lack the specificity necessary for guiding researchers in any substantive sense: they impose restrictions on the theories-to-be-constructed that still leave too much leeway to significantly help MONDian researchers in building a theory. The complexity of extensions of MOND, and the variety they display amongst them illustrate this. Both features vitiate any claims to an autonomous evolution of MONDian theories: it's difficult \textit{not} to charaterise it in terms of  ``pedestrian trial-and-error"---for Lakatos, the badge of immaturity. In short, the MONDian principles lack the heuristic power, characteristic of research programmes as Lakatos envisions them.  
    
    But suppose that one were to regard MONDian theories as a research programme sui generis. Should we, on Lakatosian standards, pledge allegiance to it---and drop GR's research programme with its Dark Matter postulate?\footnote{We don't take Lakatos---unlike, for instance, Feyerabend---to endorse a whole-hearted pluralism about research programmes. Nonetheless, he shows awareness of the salubrious effects of heterodox \textit{minority} views: ``(n)evertheless there is something to be said for at least some people sticking to a research programme until it reaches its 'saturation point'; a new programme is then challenged to account for the full success of the old.” \textcite[p.69]{Lakatos1978-LAKTMO}}
    The absence of a powerful---or, in fact, \textit{any} substantive---positive heuristic speaks against this: although degeneracy typically occurs, ``when the positive heuristic ran out of steam" \parencite[p.52]{Lakatos1978-LAKTMO}, it's clear that for Lakatos, a feeble positive heuristic bodes ill for a research programme. 
    But also Lakatos' other criteria for elimination of a research programme in favour of another aren't satisfied. 
    
    \item Contra Merritt, MONDian research \textit{hasn't} evolved progressively! MOND's relativistic extensions have been designed, through motley trial-and-error tinkering, to ensure compatibility with the available data. Whatever consequences these theories entail, they haven't been confirmed: in fact, they don't anticipate novel facts that have subsequently been corroborated.\footnote{Even \textcite[p.182]{Merritt2020} admits as much!} In fact, the standard relativistic MONDian theory, TeVeS, has been refuted (more on this in §\ref{teves}). \textcite[p.13]{Bekenstein:2006bya}, for instance, admits: ``relativistic MOND as here described has developed from the ground up, rather than coming down from the sky: phenomenology, rather than pure theoretical ideas, has been the main driver. Actually a large industry flourishes on the sidelines with imaginative ideas from first principles regarding the essence of MOND. I have not touched here on these motley approaches because they have given so little that is observationally viable".

    We are thus left with a research programme, made up of only two stages, Milgrom's original phenomenological formula and non-relativistic theories.\footnote{\textcite[Ch.7]{Merritt2020} mentions another stage of the alleged MONDian research programme, Berezhiani and Khoury’s superfluid dark matter theory (see e.g.\ \textcite{Khoury:2021tvy} for a recent review). But first, as Merritt himself admits (p.182, fn.1), it belongs rather to the standard cosmological model's research programme---a proposal for Dark Matter. (Cf.\, however, \textcite{Martens2020-MARCOT-41, Martens2020-MARDM-13}, who question the Dark Matter/modified gravity dichotomy). Secondly, Merritt also admits that (p.203), it makes ``at best a modest contribution to the progressivity of the Milgromian research program. [...] (I)t remains unclear whether [the theory] is an empirically progressive step" (ibid.) (see also \textcite[Sect.5]{Massimi:2018hzy}).} It's one thing to grant these theories explanatory and predictive successes in their intended domain (as expounded by \textcite{Merritt2020}); but it would surely be far-fetched to elevate this pair to the status of a research programme! Not only because a research programme proper arguably needs more than two stages does this strike us as implausible: first, MOND barely counts as a theory in its own right (as even admitted by many of its proponents, see §\ref{mond}), and secondly, the second stage is comprised of a \textit{non-relativistic} theory---which as such simply can't compete with the standard research programme, based on GR and Dark Matter. 
    \item Intending to promote MONDian research as superior to the standard model, \textcite{Merritt2020} has, with MSRP, chosen the wrong tool: even if his Lakatosian analysis were convincing, the Lakatosian framework can't achieve what he solicits.\footnote{This shortcoming---if it is indeed one---is an instances of precisely what Feyerabend attacks in Lakatos (see \textcite[p.7]{Barseghyan2017-BARHCA-4}). As \textcite[p.77]{Laudan1977-LAUPAI} bluntly puts it: ``[Lakatos] cannot translate his assessments of progress into recommendations about cognitive acion."}
    
    Lakatos keeps the appraisal of a research programme separate from questions of heuristic counsel. Lakatos confines his methodology to the former task. Eo ipso, such theory appraisal remains retrospective: it can at most deliver an assessment of a research programme's status quo---how it has performed to-date. This, however, entails nothing about the likelihood of \textit{future} performance. Lakatos desists from any ambitions for prospective theory evaluations, and any methodological guidance, predicated on this.
    
    Moreover, he stresses that his MSRP can only be meaningfully applied with \textit{considerable} hindsight---with due temporal distance (cf.\ \textcite[p.1]{DeBaerdemaeker2020-DEBJSS}): ``(i)t takes a long time to appraise a research programme: Minerva's owl flies at dusk." \parencite[p.149]{Lakatos1978-LAKTMO}. The objects proper for which MSRP is supposed to be a touchstone are episodes in the \textit{history} of science. MSRP ``does not offer instant rationality. One must treat budding programmes leniently. Programmes may take decades before they get off the ground and become empirically progressive" (op.cit., p.6). By Lakatos' own lights, he would have to decline to pass judgement on MONDian research as too young an area of research: ``(o)nly an extremely difficult and---indefinitely---long process can establish a research programme as superseding its rival" (op.cit., p.76). Advocates of MONDian research may---on Lakatosian standards, legitimately---request a grace period before their research agenda is dismissed.\footnote{This request strikes us as perfectly adequate---and as reflecting the already established practice in the scientific community: albeit a small research community, its work is being discussed to an extent that at least matches the attention devoted to other marginal research communities.} Yet, it's clear that devolving the methodological assessment of the acceptability of MONDian research upon posterity in an indefinite future dashes the hopes of those who, like us, think that philosophy of science can and should bring something to the table of \textit{contemporary} debates in physics. 

\end{itemize}

\subsection{Laudan---An Effectively Problem-Solving Research Tradition?}\label{laudan}

We’ll conclude our survey of the classical methodologies’ appearance in the MOND-debate with \textcite{Laudan1977-LAUPAI, Laudan1996-LAUBPA-3}. In the MOND-related literature, he plays only a peripheral role.\footnote{While we’ll here answer the question that \textcite{Martens:2022dsg} ask regarding the extent to which the MOND/Dark Matter debate ought to be construed in terms of Laudanian research traditions, to our knowledge, neither positive nor negative claims to that effect are found in the literature. The reference that Martens et al. give, \textcite{pittphilsci19913}, doesn’t deal with this issue (nor do any of Merritt’s earlier works); Merritt’s invocations of Laudan are primarily concerned with different notions of predictive novelty.}  Yet, a discussion of some of his ideas affords insightful perspectives: they allow a nuanced reconstruction of the scientific community’s largely negative verdict on MONDian research.

Like Lakatos, Laudan’s unit of methodological appraisal is larger than an individual theory---``research traditions”. Whereas the members of Lakatos' research programmes stand in diachronic and close, logical connections (cumulatively growing, corroborated empirical content), research traditions hang together more loosely. They are clusters of belief, or frameworks; their member theories---some of them ``contemporaneous”, others ``temporal successors of earlier ones” \parencite[p.78]{Laudan1977-LAUPAI}---stand in relations of family resemblance. ``Generally, [research traditions] consist of at least two components: (1) a set of beliefs about what sorts of entities and processes make up the domain of inquiry; and (2) a set of epistemic and methodological norms about how the domain is to be investigated, how theories are to be tested, how data are to be collected, and the like" \parencite[p.83]{Laudan1996-LAUBPA-3}. 

The function of a research tradition is to circumscribe ``a set of guidelines for the development of specific theories” \parencite[p.79]{Laudan1977-LAUPAI}, ``to provide us with the crucial tools we need for solving problems, empirical and conceptual. […] the research tradition even goes so far as to define partially what the problems are, and what importance should be attached to them” (op.cit., p.82): ``they indicate what assumptions can be regarded as uncontroversial ‘background knowledge’ to all the scientists working in that tradition; […] they help to identify those portions of a theory that are in difficulty and should be modified or amended; […] they pose conceptual problems for any theory in the tradition which violates the ontological and epistemic claims of the parent tradition” \parencite[p.83]{Laudan1996-LAUBPA-3}.

Laudan’s methodology assesses theories as embedded within a research tradition. Of course, rivalling research traditions as a whole can also be compared (see \textcite[p.106]{Laudan1977-LAUPAI}); but as we’ll argue below, we won’t need that here. \textcite[p.82]{Laudan1996-LAUBPA-3} first observes that most methodologies neglect the ``much wider range of cognitive attitudes” towards theories than the “opposition between between ‘belief’ and ‘disbelief’, or more programmatically, ‘acceptance’ and ‘rejection’”. He stresses ``that there is a whole spectrum of cognitive stances that scientists can adopt towards their theories (suggested by phrases like ‘entertain’, ‘consider’, and ‘utilize as a working hypothesis’)” (op.cit., p.111). One stance, in particular, ought to be delimited from a theory’s ‘acceptance’ (or ``warranted assertibility” (op.cit., 109))---‘pursuit’ (for further details, see \textcite{Barseghyan2017-BARHCA-4}). The former is the attitude “to treat [the theory] as if it were true” \parencite[p.108]{Laudan1977-LAUPAI}; the latter amounts to deeming it worthy of further investigation, showing promise that warrants the allocation of more resources (time and effort) to elaborate it. 

It can, Laudan stresses, be rational to pursue a theory \textit{without} accepting it (and vice versa). We should, \textcite[p.82]{Laudan1996-LAUBPA-3} proposes, accept the theory with the highest problem-solving effectiveness, ``that theory which comes closest to solving the largest number of important empirical problems while generating the smallest number of significant anomalies and conceptual problems”. Contrariwise, we should pursue the most promising theories---those with the highest ``rate at which the theory has recently solved problems” \parencite[p.208]{Laudan1986-LAUSCP}. 
%``a measure of how quickly a research tradition has made whatever progress it exhibits” \parencite[p.84]{Laudan1996-LAUBPA-3} AKWARD SENTENCE. I THINK IT READS BETTER WITHOUT THROWING THIS QUOTE ON THE END.

Laudan’s framework sheds light on facets of the MOND/Dark Matter debate. We can build on our preceding (negative) analysis of MONDian research as Kuhnian paradigms and Lakatosian research programmes. MONDian theories are best viewed as modifications of GR that recover salient MONDian traits in a suitable limit. They belong to \textit{GR’s} gravitational research tradition. MONDian research and GR display significant family resemblance with respect to calculational and mathematical techniques, guiding physical and formal principles (e.g.\ general covariance, causality principles), interpretative-ontological guidelines (gravity being represented at least partially via spacetime geometry), and agreement on relevant empirical phenomena in need of explanation.\footnote{Given how much of contemporary gravitational research is permeated by GR, it’s worth reminding oneself of the existence of an alternative research tradition---that of the ``particle physics approach” (see \textcite{Pitts:2017fgm, Pitts:2019sew}.)}  

A Laudanian appraisal of MONDian theories’ acceptability, however, is disheartening: their problem-solving effectiveness falls short of GR’s. True---MONDian theories and GR both grapple with empirical problems. But the overarching research tradition ranks the significance of empirical problems: this explains\footnote{\textcite[p.83]{Laudan1996-LAUBPA-3} mentions an important function of research traditions, relevant in this regard: to ``establish rules for the collection of data and for the testing of theories”.} the stronger premium on the data from cosmology, strong-gravity effects and solar system physics, and the concordance of this data (see §\ref{lcdm})---rather than on the ``messy” galactic-astrophysical data forming the bulk of the evidence in favour of MONDian theories. RMOND (to be discussed in §\ref{rmond}), the only empirically viable MONDian theory, as far as cosmology is concerned, is little studied: to what extent it’s free from empirical problems in other domains and applications (e.g.\ black holes or neutron stars or gravitational wave phenomenology) seems largely unknown. The same uncertainty besets potential conceptual problems of RMOND. Hence, it’s (at best) difficult to estimate RMOND’s acceptability, and meaningfully compare it to GR’s. 

What about pursuit-worthiness of MONDian research along Laudanian lines? Our presently limited knowledge of RMOND beyond cosmological applications compromise any estimate of the rate of progress. An estimate based on those achievements remains exiguous. To begin with, it took more than 20 years to develop a version of MOND, TeVeS, that satisfied, what the research tradition demands as empirical and conceptual sine qua nons for relativistic theories of gravity (see §\ref{teves} for details). MONDian advances in empirical problem solving were, at best, sluggish for the subsequent decade. Eventually, the main MONDian theory was refuted by cosmological evidence; at the same time, once more, the GR-based standard model was confirmed. Via a more or less artificial modification, a variant of TeVeS was devised that survived the data previously damning its precursor. We aren’t aware of any novel empirical problems that have been solved (with a potential exception in very recent times, see §\ref{teves}) and that, \textit{by the research tradition's own standards}\footnote{Thereby excluding the ``small-scale controversies" for galactic phenomena.}, are significant. The upshot is: while MONDian theorising has been making progress, its rate of progress plausibly compares poorly to that of GR.\footnote{To be fair, there hasn't been (positive) progress on the (especially particle) Dark Matter front, either---at least not empirical progress. But in comparison to MOND, standard GR's overall \textit{already demonstrated} problem-solving effectiveness still strikes us as superior (as attested to by the breath-taking range of its successful applications, see e.g.\ \textcite{ellis_maartens_maccallum_2012} or \textcite{Will:2014kxa}).} With the decades of the ``Battle of the Big Systems” in methodology \parencite{sep-rationality-historicist} over, we’ll now zoom in on a specific complaint. 

\section{Ad-hocness as a Tool for Theory Evaluation: MOND vs.\ $\Lambda$CDM}\label{adhoc}

\noindent Arguably the most common super-empirical appraisal of a theory, by scientists themselves, is the charge of ad-hocness: an ad-hoc theory is rebuked for somehow being \textit{fudged} to either conform to the evidence, or to elude refutation. In this spirit, detractors of MONDian research likewise decry it as ad-hoc. \textcite[p.14]{WallaceD} provides a representative assessment: ``[...] twenty years ago MOND was a highly plausible rival, but by now the level of contrivance and ad hoc modifications required to fit the data makes it most unlikely to be correct. And that assessment is shared by most of the astrophysics community [...]" (likewise, e.g.\ \textcite[p.320]{ellis_maartens_maccallum_2012}). How do we unpack MOND's (and its relativistic extensions') indictment as ad-hoc? And in particular, how to understand and vindicate the \textit{normative} thrust of this categorisation---the intended scathing of an \textit{epistemic} defect?

For answers, we'll adopt the (to our minds) most comprehensive and convincing account of ad-hocness to-date---Schindler's coherentist model \parencite{Schindler2018-SCHACC-4}. It will be outlined in §\ref{coherent}. We'll subsequently apply it to the main strands of MONDian research: MOND itself (§\ref{mond}), and two of its relativistic extensions, TeVeS (§\ref{teves}) and RMOND (§\ref{rmond}). The verdicts that the coherentist model delivers match, and render precise, the hunches underlying the criticism  of MONDian research, voiced by many members of the scientific community. Lastly, for a fair evaluation of MONDian research, it behooves us to compare its ad-hocness with that of the standard cosmological model (§\ref{lcdm}).

\subsection{The Coherentist Account}\label{coherent}

Consider a generic instance of Duhem's Problem, with some theory $T$ in empirical difficulties: together with background assumptions $\mathcal{B}$, $T$ conflicts with some observational data $\mathcal{O}$. By dint of some hypothesis $H$, one now tries to resolve the conflict. (We needn't specify how $H$ bails us out---that is, which element it replaces in the troublesome problem situation $T\&\mathcal{B}\&\mathcal{O}$: $H$ could be a modification of $T$, revision of some background assumptions $\mathcal{B}$, or a re-interpretation of the observational data $\mathcal{O}$.) 

Following \textcite{Schindler2018-SCHACC-4}, let's call $H$ ad-hoc, if it's introduced  \textit{arbitrarily} for the purpose of saving $T$ from empirical refutation: $H$ replaces $T$ or some of the background assumptions $\mathcal{B}$, or re-interprets the observational data $\mathcal{O}$, in a methodologically problematic way. $H$'s introduction is arbitrary---and hence the problem situation's modification via $H$ methodologically suspect---when either $H$ and $T$ don't cohere, or $H$ doesn't cohere with accepted background assumptions $\mathcal{B}$.\footnote{Such background theories (or more neutrally: hypotheses) may also include (e.g.\ interpretative) assumptions about observational data.} For our purposes, $H$ is said not to cohere with $T$ (or the background knowledge $\mathcal{B}$), if $T$ (or our background knowledge $\mathcal{B}$) gives no good reasons to believe $H$ (rather than $\neg H$).\footnote{Such a hypothesis could, for instance, also state that some variable has a particular value.} Coherence presupposes (logical) compatibility; but coherence proper goes beyond that. ``Good reasons" are supposed to be construed liberally. First and foremost, they include (but aren't necessarily limited to\footnote{An example of a \textit{non-explanatory} modification of, say, Maxwellian electrodynamics that would prima facie still count as coherent with it is the inclusion of a mass term (see e.g.\ \textcite{Pitts:2015bja}). We take this kind of coherence to be an instance of what \textcite[p.101]{Zahar1973-ZAHWDE} may have had in mind with ``a modification of the auxiliary hypotheses which (accords) with the spirit of the heuristic of the programme".}) a broad range of explanatory relations (e.g.\ deductive-nomological, causal, unificationist, or structural): a good reason for $H$ is typically an answer to a ``why-$H$ question" (see e.g.\ \textcite[Ch.5]{VanFraassenBas1980-VANTSI}).

It's vital not to mistake eschewal of ad-hocness for an infallible guide to truth. A few remarks are therefore in order on the relationship between ad-hocness and truth. First, plausibly, at least a pinch of ad-hocness seems ineluctable: ``[...] one might question whether there is any prominent scientific theory which does entirely without ad hoc assumptions" \parencite[p.136]{Schindler2018-SCHACC-4}. Some things just ``don't hang together": they just happen to be what they are---contingently given. As we'll see below, a further reason for this omnipresence of ad-hocness lies in the graded nature of ad-hocness. Secondly, established background theories and assumptions can, and typically do, change. A hypothesis erstwhile acquitted of ad-hocness can thus appear in a new light: formerly good reasons for it may no longer hold. They may even have been ousted by reasons \textit{against} it. The opposite is, of course, likewise possible. Thirdly, ad-hocness is neither necessary for a hypothesis' falseness---nor is \textit{lack} of ad-hocness sufficient for its truth. History teems with examples of ad-hoc hypotheses that were subsequently \textit{corroborated} (e.g.\ Pauli's neutrino hypothesis or Planck's postulate of the elementary quantum of action); conversely also (relatively) non-ad hoc hypotheses turned out to be wrong (e.g.\ Leverrier's postulate of the planet Vulcan to account for the anomalous precession of Mercury). 

If (non-)ad-hocness and truth stand in no straightforward relation, what is supposed to be objectionable about ad-hocness? What is its normative dimension grounded in? Deprecation of ad-hocness, on the coherentist model, is derived from a premium on epistemic conservativism---as a makeshift rule of thumb for theory appraisal \textit{faute de mieux}: we decide to value hypotheses with little ad-hocness in order to ``piggy-back" as much as possible on the epistemological warrant of the theory in empirical trouble or the background assumptions.\footnote{We'll set aside the potential epistemological significance of coherence as a theory virtue, see e.g.\ \textcite[p.165]{Putnam1975-PUTTMO}.} Neither $T$ nor $\mathcal B$ can obviously lay claim to apodeictic certainty. With respect to its conduciveness to truth, avoidance of ad-hocness can only be as good as our trust in the theory saved by the ad-hoc hypothesis, and our background knowledge. Judgements of a hypothesis' ad-hocness thus gauge its plausibility, its consonance with background \textit{belief}---rather than its likelihood to be \textit{true}\footnote{Herein we deviate from Schindler's own position: for him, avoidance of ad-hocness is a propitious strategy for truth. To our minds, for the present context plausibility (i.e.\ coherence with background knowledge) seems a more realistic, modest goal. \textcite[p.43]{Massimi2014-MASWAD} rightly remark about the astrophysics and cosmology in question: ``the basic laws of physics have been obtained entirely from experiments carried out on Earth [...], so it is a considerable extrapolation to assume that no new physics exists that only manifests itself on very large scales. Nevertheless, this is how cosmologists choose to play the game: otherwise, it is impossible to make any predictions for cosmological observations." We thank Schindler for discussion.}---in the absence of more compelling (in particular: empirical) reasons; the methodological rule to minimise ad-hocness in theory choice reflects the need (or penchant) to veer as little as possible from well-established knowledge.\footnote{This is compatible with drastic revisions of that knowledge, or even scientific revolutions---provided the empirical or theoretical reasons for such innovation are sufficiently strong.}       

The coherentist account of ad-hocness overcomes most of the shortcomings of alternative accounts; hence its application to MOND is particularly germane. In several respects, the account stands out (see \textcite[Ch.5]{schindler_2018} for details):
\begin{itemize}

\item Despite subjective components (see below), overall it's objective: whether coherence relations obtain is an objective matter. Objectivity is a natural desideratum: diagnoses of ad-hocness are supposed to expose an epistemic defect. A non-objective account would enfeeble the force of such a diagnosis.

\item The account carries manifest epistemic weight: it flags a hypothesis' epistemic defect. Insofar as we expect our knowledge to cohere, lack of coherence is to be shunned. The account derives its normative force from a natural epistemic desideratum of coherence---the availability of reasons.

\item The account allows for gradations: rather than a binary property, ad-hocness comes in degrees. They range from maximal ad-hocness (tantamount to inconsistency with background knowledge or good reasons \textit{against} $H$) to non-adhocness (i.e. $H$'s strict derivability from $T$ and $\mathcal B$). Most realistic examples lie on the spectrum between these extremes. Moreover, coherence itself admits of degrees: good reasons for a hypothesis may differ in strength and quality. Furthermore, since different coherence relations between the $H$ and $T$ or $\mathcal B$ are possible, they may be assigned different weights/importance. Reminiscent of the Kuhnian weighting problem (§\ref{kuhn}), this is arguably a subjective matter, left to the researcher's individual discretion. Altogether, the graded nature of ad-hocness can be viewed as a benefit. The coherentist account thereby gains flexibility---another desideratum for a realistic account: ``degrees of ad hocness allow us to say one theory ought to be preferred over another, regarding ad hocness, even if neither theory manages entirely without ad hoc hypotheses: one theory might just invoke fewer ad hoc hypotheses than the other" \parencite[p.13]{Schindler2018-SCHACC-4}.

\item The coherentist account yields verdicts that mesh with many historical examples, typically judged---by both contemporaneous and modern scientists (as well as philosophers)---as ad-hoc. This kind of extensional adequacy is a plausible meta-methodological desideratum (\textcite{Schindler2013-SCHTKM}; \textcite[Ch.7]{schindler_2018}). 

\item The coherentist account explains and, to some extent, subsumes the merits and intuitive appeal of rivalling accounts. Consider, for instance, the model of ad-hocness given by \textcite{WORRALL201454} in terms of a hypothesis' free parameters, particularly relevant for our discussion of MOND's ad-hocness (see also \textcite[Sect.3]{MartensKing}). Worrall's verdict on the methodological reprehensibility of ad-hocness thus defined seems to rest on (Popperian) qualms about such a hypothesis' susceptibility to independent tests (see e.g.\ \textcite[Ch.3]{schindler_2018}). Recovering the intuition that, ceteris paribus, a theory with few parameters should be preferred to one with many, the coherentist account offers a different rationale. For each parameter one may wonder why it takes one particular value rather than another. Sometimes, we won't obtain an answer, of course: some things must be accepted as brute facts (cf.\ \textcite{Hossenfelder2019-HOSSFE, Baras2022-BARCFE-5}). But---the more options there are in the theory's parameter space, the more pressing the question becomes: we covet good reasons for any particular choice. 

\end{itemize} 
%With these tools in hand, let's see how they play out in the developments of MONDian research.

\subsection{MOND}\label{mond}

How does MOND fare on the coherentist account of ad-hocness?\footnote{The coherentist model doesn’t invariably favour conservativism with regards to \textit{laws}, over conservativism with regards to \textit{entities}. Otherwise, the coherentist model would seem \textit{ab initio} prejudiced towards Dark Matter; such a bias would subvert the model's suitability for a fair assessment of a modified law of gravity as an alternative Dark Matter. Background beliefs include both laws as well as putative entities (e.g.\ as the entities to whose existence the theories are committed). Typically, it will be delicate to non-arbitrarily weigh coherence with respect to one’s background knowledge about laws against coherence with respect to entities (cf.\ also \textcite{Vanderburgh2014-VANQPE}). But this problem is characteristic for weighing the various coherence relations in which a hypothesis can stand with respect to background knowledge (see also \textcite[p.136]{schindler_2018}). We thank an anonymous referee for pressing us on this interesting point!} It confirms ``an argument often directed against MOND that, as a theory, it is ad hoc and incomplete" \parencite[p.1]{Sanders:1997we}. 

Begin with coherence with background \textit{empirical facts}. In this regard, the coherentist account yields an at first blush favourable verdict on MOND. As even sceptics acknowledge (e.g.\ \textcite{Aguirre} or \textcite[p.7]{pittphilsci18810}), MOND splendidly systematises galaxy-scale astrophysical phenomenology. In a thought-economic manner---in particular, without having to postulate Dark Matter, baryonic or non-baryonic---it accurately summarises sundry phenomena.\footnote{Sometimes it's suggested that MOND ``turns out to \textit{have} an intriguing connection with cosmology" \textcite[p.1, p.7, our emphasis]{Kroupa:2012pjd}, via the numerical coincidence: $a_0\approx {c H_0}/{2\pi}\approx c^2 \sqrt(\Lambda/3)/{2\pi}$, with the speed of light $c$, the Hubble constant $H_0$, and the cosmological constant $\Lambda$; also a link with Mach's Principle is hinted at. Whatever future MONDian developments might reveal, at present those connections seem coincidental. Hence it would be specious to chalk them to MOND's coherence/non-ad-hocness score.} It even enables various novel predictions; many of them have indeed been corroborated (see e.g.\ \textcite{Milgrom:2014usa}; \textcite[Ch.4]{Merritt2020}). The universality of an acceleration scale on which observable deviations from Newtonian phenomenology occur deserves to be highlighted (see e.g.\ \textcite{McGaugh:2020ppt}). By construction, MOND reproduces the familiar Newtonian phenomenology outside the deep-MOND regime. In these regards, MOND accomplishes astonishing coherence amongst a body of knowledge of empirical/observational facts.        

This positive impression must be qualified, however. Two instances of MOND's \textit{incoherence} with background factual knowledge stand out: 
\begin{itemize}
\item
MOND is a modification of Newtonian gravity in the low-acceleration limit. Therefore, MOND doesn't cohere---and rather conflicts---with post-Newtonian (and, a fortiori, \textit{fully} general-relativistic) data. For instance, it can't account for gravitational lensing, or Mercury's perihelion precession---nor the relativistic phenomenology proper, such as effects related to gravitational radiation. In Popperian terms, this failure to cover nigh-universally recognised empirical facts counts as a falsification (§\ref{popper}); it certainly generates significant ad-hocness in terms of coherence with established background empirical knowledge.
\item
Likewise, the so-called ``external field effect" compromises MOND's compatibility with background knowledge of empirical facts. It's ``not a prediction but a phenomenological requirement [...]. In his original papers Milgrom noted that open star clusters in the Galaxy do not show evidence for mass discrepancies, even though the internal accelerations are typically below $a_0$. He therefore postulated that the external acceleration field of the Galaxy must have an effect upon the internal dynamics of a star cluster" \parencite[p.271]{Sanders:2002pf}. The data from the observed of star clusters in the Milky Way ``(led) Milgrom to the realization of the external field effect; that is to say, it is the total acceleration, internal plus external, that must be included in [the above Eq. xxx]" \parencite[p.10]{Sanders:2014xta}. This modification certainly counts as an adjustment in light of an empirical anomaly. Its induced ad-hocness is arguably mild, however, given MOND's non-linear framework (ibid.). 

More delicate is the external field effect's conflict with the Strong Equivalence Principle. According to the latter, a self-gravitating system's internal dynamics in free-fall in an external gravitational field doesn't depend on the external field strength. The Strong Equivalence Principle is a cornerstone of GR, and valid in good approximation (see e.g.\ \textcite{Will:2014kxa}). On the one hand, the external field effect thus induces incoherence with respect to our \textit{theoretical} background knowledge. On the other hand, according to \textcite{Chae:2020omu}, rotationally supported galaxies seem to display a robust violation of the Strong Equivalence Principle in the deep-MOND regime. Should Chae et al.'s results live up to critical scrutiny, MOND's incoherence with our theoretical background knowledge---and hence MOND's ad-hocness in this regard---will turn out to be overridden by novel \textit{empirical-factual} background knowledge. 
\end{itemize}
Let's dwell on coherence with \textit{theoretical} background knowledge.\footnote{Following \textcite[p.112]{sanders_2016}, we construe the ``incompleteness", which Sanders referred to in the above quote, as absence of a relativistic theory (more on this in §\ref{teves} and §\ref{rmond}). ``Without a relativistic theory, it is not possible to address cosmology and the formation of structure---those large-scale issues in which the standard paradigm does so well" (ibid.). This ``incompleteness" entails incoherence with respect to both empirical-factual and theoretical background knowledge.} Here, MOND receives a devastatingly poor rating.

\begin{itemize}
\item 
One may be surprised by MOND's dependence on a preferred acceleration scale: all other fundamental forces, by contradistinction, are \textit{distance}-dependent. Hence, one wouldn't expect MONDian gravity to introduce an \textit{acceleration}-dependent scale. This form of ad-hocness is, we think, perhaps not particularly problematic: why \textit{shouldn't} gravity be special also with respect to its \textit{scale}-dependence? 
\item 
The interpolation function $\mu$, a free function in MOND, is a thornier point. A free parameter diminishes a theory's coherence, also for the coherentist model. A fortiori, a free \textit{function}---tantamount to a nondenumerable \textit{infinity} of free parameters---does! For many applications $\mu$'s specific form is admittedly irrelevant, \textit{provided} that it has the right asymptotic behaviour.\footnote{Not all forms of $\mu$ in the transition region are consistent with solar system data: some choices therefore \textit{are} detectable, see \textcite[p.65, fn.10]{Merritt2020}.} Even so, this very asymptotic behaviour--MONDian behaviour in the deep-MOND regime, and Newtonian behaviour outside of it---itself detracts from MOND's coherence: why this particular asymptotic behaviour, rather than some other?
\item
MOND's coherence with \textit{non}-gravitational physics is, at best, questionable. \textcite[p.3]{Milgrom:2014usa} concedes: ``(w)e do not know to what extent and how MOND affects nongravitational phenomena such as electromagnetism (EM). For example, if there is a consistent way to extend and apply the basic tenets to nongravitational physics."
\item 
MOND violates linear momentum conservation \parencite{1984ApJ...286....3F}: generically, an isolated self-gravitating system experiences non-zero total acceleration. Conservation of momentum, however, is deeply entrenched in physics. In fact, it's intimately linked with the homogeneity of space itself. By the same token, MOND violates conservation of angular momentum and energy. MOND thus \textit{conflicts} with a high-level, physical principles of our background knowledge; the resulting ad-hocness is severe.\footnote{One might argue that in GR those conservation principles must be revised or even abandoned (see e.g.\ \textcite{DuerrThesis,Hoefer2000-HOEECI-2}); hence, MOND's violation of them might be perceived as little problematic. But this overlooks two differences. First, whether GR mandates such revisions is controversial (see e.g.\ \textcite{DeHaro:2021gdv}, \textcite{Pitts:2009km}); by contrast, one can \textit{demonstrate} that MOND violates, say, conservation of linear momentum. Secondly, if GR indeed mandates such revisions, they are under-girded by a fully-developed, empirically well-corroborated general theory: GR's successes, empirical and theoretical, are so overwhelming that, in case of conflict, modifying conservation principles seems the lesser evil. That \textit{wouldn't} be the case for MOND.} 
\item 
\textcite{1984ApJ...286....3F} points out another, related problem: for MOND, the centre-of-mass theorem fails. In contrast to Newtonian Gravity, an isolated system's total momentum no longer coincides with that of a particle with the system's total mass, moving with the velocity of the center of mass. As a result, we can't treat the dynamics of multi-particle systems, such as galaxies, effectively as a one-particle system. But it's \textit{multi-particle systems} that astronomical data refer to. He concludes: either MOND's core equation generically doesn't apply to galaxies---or generically, it doesn't apply to individual particles. Either way, MOND's dynamics ``is therefore incomplete. [...] At present, Milgrom's law must be thought of a phenomenological modification of Kepler's laws rather than a systematic modification of Newtonian dynamics" (op.cit., p.4). This can be explicated as a charge of coherentist ad-hocness: MOND is a specific, low-level regularity that either conflicts with the more general established framework theory in our background knowledge, or that forces us to discard this framework---without offering an alternative in its place. Either way, MOND comes out as severely ad-hoc.
\end{itemize}

In conclusion, MOND doesn't square with our background knowledge in several regards. Our \textit{theoretical} background knowledge flat-out contradicts it. MOND coheres only with \textit {some}\footnote{In this respect, it's worth reiterating that at present, it would be premature to claim that only MOND, but \textit{not} the standard model, can do justice to this phenomenology, see §\ref{lcdm}}---but not all---of our empirical-\textit{factual} background knowledge, where again some of our empirical background knowledge contradicts it as well. On the coherentist account, MOND comes out as strongly ad-hoc. 

The verdict on MOND's ad-hocness chimes with widespread opinion in the physics literature.\footnote{In this vein, \textcite[p.581]{2006eac..book.....S} rejects MOND as incoherent with theoretical and empirical background knowledge.} Themselves advocates of MONDian research, \textcite[p.297]{Sanders:2002pf} for instance admit: ``(i)n spite of its empirical success, MOND remains a largely ad hoc modification of Newtonian gravity or dynamics without connnection to a more familiar theoretical framework. [...] The original algorithm [...] cannot be considered as a theory as a successful phenomenological scheme [...]".

Some of MOND's theoretical shortcomings have been amended in subsequent MONDian theories. To some extent, they mitigate the ad-hocness arising from the shortcomings pointed out above. With respect to ad-hocness simpliciter, however, such theories are ambiguous. Despite interesting theoretical features (some of which might even count as successful predictions, see e.g.\ \textcite[Ch.2]{Merritt2020}), they tend to engender new sources of ad-hocness themselves---both with respect to our received (primarily theoretical) background knowledge, as well as in terms of those theories' lack of coherence with Milgrom's original proposal. Here, we won't analyse those non-relativistic successor theories of MOND (such as in those in \textcite{Bekenstein:1984tv} or \textcite{Milgrom:2009ee}). Instead, we'll move on directly to MOND's relativistic field-theoretical extensions; only they stand a chance to ``jump through the loops" for ``a mature theory" \parencite{Carrollblog}.         

\subsection{TeVeS}\label{teves}

Ad-hocness (construed in the coherentist way) paved the way towards viable relativistic versions of MOND. Early attempts  (see e.g.\ \textcite{Bekenstein:2005nv}; \textcite{Bekenstein:2004ne} for historical reviews) failed to cohere with general field-theoretic and/or relativistic principles (such as causality or absence of preferred frames). Struggling to adequately replicate other successes of GR (such as gravitational lensing), they also failed to cohere with firm empirical-factual background knowledge. 

Eventually, TeVeS emerged as the first satisfactory relativistic version of MOND (see e.g.\ \textcite{2009CQGra..26n3001S} or \textcite{Bekenstein:2011ojw} for reviews); until the detection of the binary neutron star merger event in 2017 (see below), it remained the relativistic flagship of MONDian research. TeVeS represents gravity by a metric tensor $g_{\mu \nu}$, a (dual) vector field $\mathcal U_\alpha$, and a scalar field $\phi$; hence TeVeS's appellation---``\textit{Te}nsor-\textit{Ve}ctor-\textit{S}calar Gravity". All three combine to define TeVeS's so-called ``Bekenstein metric": $\tilde{g}_{\mu \nu}: =e^{-2 \phi} g_{\mu \nu}-2 U_{\mu} U_{\nu} \sinh (2 \phi)$. Matter degrees of freedom couple to this metric rather than to the ``gravitational" metric $g_{\mu \nu}$ in the standard, general-relativistic manner (viz.\ universally, see e.g.\ \textcite[Ch.9.5]{RBrown2005-RBRPRS-2}). 

We obtain TeVeS's dynamics for these variables from the schematic\footnote{The details needn't detain us, see e.g.\ \textcite{Bekenstein:2004ne}, also for the (lengthy) and cumbersome equations of motions, resulting from variation.} action \parencite{Bekenstein:2004ne}
\begin{equation}
    S_{TeVeS} = \int d^4x \left[\mathcal{L}_{EH} (g_{\mu\nu})+ \mathcal{L}_S (\phi, \sigma, g_{\mu\nu}, F) + \mathcal{L}_V (U_\mu, g_{\mu\nu}) + \mathcal{L}_m (\tilde{g}_{\mu\nu}, \psi)\right].
\end{equation}
Its components comprise GR's standard Einstein-Hilbert Lagrangian $\mathcal{L}_{EH}$, a Lagrangian for scalar fields $\mathcal{L}_S$, a Lagrangian for a vector field $\mathcal{L}_V$, and the matter Lagrangian $\mathcal{L}_m$ for generic matter fields $\psi$. 

Three comments are in order regarding the composition of the actions for the scalar and vector, respectively. 
First, the Einstein-Hilbert part is defined via $g_{\mu \nu}$. The latter also figures in the scalar and vector parts. By contradistinction, the matter action features the \textit{Bekenstein metric}, $\tilde{g}_{\mu \nu}$. As a result, test-particles follow geodesics of the Bekenstein metric $\tilde{g}_{\mu \nu}$---not those of $g_{\mu \nu}$.\footnote{In the evocative terminology of \textcite{RBrown2005-RBRPRS-2}, rods and clocks ``monitor" (or ``advert to") the Bekenstein metric (see Brown's explicit discussion of TeVeS on p.172).}   
Secondly, the scalar part of the action, $\mathcal{S}_S$, in addition to a more familiar dynamical scalar field $\phi$, contains a \textit{non}-dynamical, auxiliary variable, the dimensionless scalar field $\sigma$, as well as an arbitrary function, $F(\sigma)$. The latter is subject to the constraint that the Newtonian or MONDian limit be recovered in quasi-static solutions. $\mathcal{S}_S$ is meticulously constructed to forestall superluminal propagation of perturbations of $\phi$. (This constraint still leaves much leeway; $F$ remains empirically significantly underdetermined---a source of ad-hocness, see below.)
$\mathcal{S}_S$ requires a new coupling constant (typically chosen as a scale parameter). Thirdly, the action for the vector field consists of a Maxwellian term (mandating a new, dimensionless coupling constant for the vector) and an additional term with a Lagrange multiplier that ensures that the vector field remains time-like.

From this outline of TeVeS, we can glean its core forms of ad-hocness. We begin with TeVeS's achievements. It's explicitly constructed in conformity with standard field-theoretic desiderata. This removes a major source of ad-hocness (incompatibility with background theoretical knowledge) that afflicted other (non-relativistic and relativistic) extensions of MOND. TeVeS indeed accounts for much---but as we'll see below, \textit{not} all---of the phenomenology that GR accounts for. In particular, TeVeS is consistent with solar-system tests (for suitable choices of TeVeS's parameters). Importantly, in a suitable limit, TeVeS recovers MONDian phenomenology.\footnote{We gloss over subtleties that can arise in systems that aren't highly symmetrical. For them, MOND \textit{can} deviate from TeVeS's quasi-static limit (see \textcite[Sect.2.5]{2009CQGra..26n3001S}).} In this sense, it achieves coherence---and hence non-ad-hocness---with respect to the pertinent empirical-factual background knowledge, as well as, to some extent, the MOND-related background theoretical knowledge.\footnote{NB: ``It should be stressed that the action for the scalar field [i.e. $\mathcal{S}_S$] has been constructed such that the theory has a MONDian non-relativistic limit, under the right conditions, for specific choices of functions [$V(\sigma$] [...]" \parencite[p.53]{Clifton:2011jh}.} 

The construction of such a theory is nothing short of a technical feat of wizardry. TeVeS's achievements, however, are purchased by grave ad-hoc liabilities. We group them into three broad categories.
\begin{enumerate}
    
    \item \textit{Proliferation of parameters and structure}:
    
    As discussed in §\ref{coherent} and §\ref{mond}, the number of free parameters (and, a fortiori, free functions) abounds to coherentist ad-hocness. In the guise of $F$, a free function, TeVeS inherits a counterpart of the interpolation function $\mu$ from its MONDian origins. 
    
    \textit{In addition} to $F$, TeVeS introduces three\footnote{Spherically-symmetric solutions of the original TeVeS in fact turn out to suffer from instabilities. To cure this defect, \textcite{skordis2008} proposed a generalisation of TeVeS. It introduces \textit{four} additional free parameters in the vector Lagrangian. Instabilities can take a variety of forms in modified gravity theories. See \textcite{Wolf:2019hzy} and \textcite{Rubakov:2014jja} for some further discussions on instabilities that can arise in the modified gravity and cosmological context.} new free parameters, the coupling constants for TeVeS's additional gravitational fields (the scalars $\phi$, and $\sigma$ and the vector $A_\alpha$, respectively), which themselves represent multiple additional new gravitational degrees of freedom. On this dimension of ad-hocness, TeVeS thus fares even worse than MOND.
    
    TeVeS's peculiar structure is a kindred source of ad-hocness. We'll call it ``\textit{structural} ad-hocness". Already ``(t)he existence of a scalar \textit{and} a vector field may seem odd at first [...] \parencite[p.52, our emphasis]{Clifton:2011jh}. TeVeS's oddities are indeed manifold. For instance, it relies on a non-dynamical scalar, $\sigma$, in its action for the gravitational scalar. The latter has an unusual form, as compared to other theories with scalars (e.g.\ the Klein-Gordon field) and/or vectors (e.g.\ Yang-Mills theories). It's engineered to evade causal anomalies. By the same token, the Lagrange multiplier in the gravitational vector's action smacks of artificiality: its sole purpose is to enforce the vector's time-likeness and normalisation. Why impose these constraints? No reasons seem forthcoming other than incorporating a ``phenomenological requirement for the theory to give the right bending of light" \parencite[p.5]{2009CQGra..26n3001S}. The same applies to the construction of the Bekenstein metric, employed in the matter action: the only reason for its introduction and specific, intricate form is to skirt immediate empirical refutation. 
    
    In sum, vis-à-vis other theories, TeVeS's mathematical structure is rather abnormal. Insofar as its only justification consists in escaping conceptual or empirical disaster, it may plausibly be viewed as contrived.
    Are \textit{sufficiently persuasive} reasons forthcoming for TeVeS's fine-tuning at the structural level? If one disputes this (as we're inclined to), TeVeS doesn't cohere \textit{structurally} with our theoretical background knowledge.  

\item 
\textit{Gravitational Waves}:

According to TeVeS, gravitational and electromagnetic waves don't travel at the same speed. A binary neutron star merger event recently allowed to test this prediction: ``(t)he coincident observation of gravitational waves and electromagnetic radiation from GW17081781 has allowed to set very stringent constraints on the propagation velocity of gravitational waves. The fact that it does not differ from the speed of light by more than one part in $10^{15}$ severely constrains all theories of modified gravity in which gravitational waves travel on different geodesics with respect to photons and neutrinos. This has in particular allowed to rule out Bekenstein’s Tensor-Vector-Scalar (TeVeS) theory" \parencite[p.5]{Bertone:2018krk}. While a \textit{further modification} of TeVeS can accommodate these findings, TeVeS as it stands fails to cohere with empirical-factual background knowledge and is indeed refuted by it (see \textcite{2022FoPh...52...31A} for a detailed philosophical analysis).
    
\item 
\textit{Dark Matter}: 

TeVeS cannot account for important empirical facts without introducing Dark Matter species. Empirical adequacy thus necessitates TeVeS's ad-hocness with respect to basic MONDian and, in some cases, particle-physics background assumptions.
    
        \begin{itemize}
            \item TeVeS doesn't produce enough gravitational lensing to account for observations in galaxy clusters. Missing mass is needed to make up for the discrepancy. One common riposte in the MOND community postulates very massive neutrinos to  fill the gap \parencite{10.1046/j.1365-8711.2003.06596.x}. This option has plausibly been ruled out \parencite{Mertens:2016ihw}; it doesn't cohere with established background knowledge (from particle physics).
            
            \item TeVeS violently disagrees with the Matter Power Spectrum (MPS), the measured density contrasts as a function of scale. In a universe dominated by baryonic matter, so-called baryonic acoustic oscillations, perturbations in the density of visible matter, would undulate. The presence of large amounts of non-baryonic Dark Matter would smooth out the MPS on the measured scales; gravitational potential wells created by Dark Matter would suppress the density perturbations. Observations of the cosmic microwave background confirm this: the measured MPS indeed corresponds to the patterns expected in a universe dominated by non-baryonic Dark Matter \parencite{Dodelson:2011qv}. Empirical adequacy thus requires the introduction of these additional matter components. 
            
            Allowing for such ``dark", non-baryonic components, however, doesn't cohere with \textit{MONDian background principles}: ``that clusters contain non-Baryonic cold [Dark Matter] between the galaxies, while logically possible, seems hardly justifiable in view of MOND's overall philosophy" \parencite[p.4]{Bekenstein:2006bya}.  
            \item Unlike the standard model, TeVeS cannot produce the CMB (a claim to be somewhat qualified with respect to a delicate detail, to be discussed shortly). Its power spectrum, taken in its entirety, comes out strongly against TeVeS. Here again, TeVeS doesn't cohere with widely accepted background empirical-factual knowledge.  
    
            The amplitudes of CMB peaks are exquisitely sensitive to the early universe's matter density. As \textcite[p.152]{Merritt2020} reports, TeVeS appeared to \textit{correctly predict} the ratio of the first and second CMB peaks. Initially, this number was problematic for $\Lambda$CMD cosmologists, who expected a lower number due to the nuclear physics constraints on the baryonic matter density in the early universe that were available \textit{at that point in time}. However, Merritt himself acknowledges later on (op.cit., p.178) that this prediction follows from the conjunction of using these old constraints for the baryonic matter density, and assuming that there is no cold Dark Matter. Hence, it would be misleading (contra Merritt) to declare it a prediction of TeVeS (or MOND) \textit{per se}. Rather, it's the ratio one would expect in \textit{any} FLRW universe with that particular baryonic matter density and no other matter species present.

            Those early constraints have subsequently been revised; the baryonic density was corrected upwards and the ratio of the first and second peaks now match what we would expect from the $\Lambda$CDM universe. Insofar as the $\Lambda$CDM model incurs some (coherentist) ad-hocness, we concur with Merritt (op.cit., p.177). Insofar as he repudiates it, however, as ``an \textit{unreasonable} modification", we dissent; four reasons alleviate the ad-hocness of the correction.\footnote{This isn't to deny some lingering issues with inferring primordial abundances from stellar observations, and particularly with lithium. But ``nature is entitled to hide baryons" \parencite[p.338]{Peebles2020}: there are many poorly understood processes likely to affect such observations; they can plausibly hide significant quantities of baryons that we would otherwise infer from stellar processes. This is surely an interesting anomaly that does not yet have a full resolution. But we should be careful to consider the whole CMB picture before throwing out $\Lambda$CDM (cf.\ \textcite[Sect.3.3]{DeBaerdemaeker2020-DEBJSS}). The totality of evidence indicates agreement with the $\Lambda$CDM model (see also §\ref{lcdm}).}
             
            First, in general, CMB constraints are considered far more reliable than other means of determination: other methods of deriving primordial abundances (whether from Big Bang Baryogenesis or late-time observations) rely on physics that is far more complicated and less well-understood, whereas CMB estimates all come from comparatively simple, well-understood linear processes \parencite[p.448]{2006eac..book.....S}. Secondly, subsequent measurements of primordial abundances, from both Big Bang Nucleosynthesis and observations of galaxy clusters, have moved closer to this value to the point that there is now a widespread convergence amongst all of these measurements \parencite{Fukugita:2004ee}. Thirdly, a universe with \textit{only} baryons makes another clear prediction: each CMB peak should be smaller than the ones anteceding it, as the oscillations in this baryon-photon fluid undergo damping. This, however, doesn't match the observations: the third peak is about even with the second peak. Hence, the CMB contains strong evidence for non-baryonic dark matter because such matter content is decoupled from the baryon-photon fluid and provides an additional gravitational potential needed to enhance the third peak to its observed value \parencite[p.114]{Famaey:2011kh}. Fourthly, and finally, the $\Lambda$CMD model generally predicted the correct shape of the CMB from the beginning of detailed numerical studies of it \parencite{PhysRevLett.72.13}; subsequent refinements have allowed it to provide a high-accuracy fit for \textit{all} of the CMB's observed features \parencite{Planck:2018vyg}. By contrast, TeVeS cannot provide an explanation for the overall picture of observed CMB features. Few options remain for advocates of TeVeS: either they have to further modify TeVeS by injecting new degrees of freedom, or they have to introduce Dark Matter \parencite[p.115]{Famaey:2011kh}. Each horn of the dilemma unleashes more ad-hocness. 
                    \end{itemize}
    
\end{enumerate}

In conclusion, TeVeS displays significant ad-hocness. It flows from four main sources: lack of structual coherence (due to TeVeS's unusual and arguably contrived mathematical structure), lack of ``external" coherence with respect to background knowledge (due to the need to postulate (baryonic) Dark Matter not warranted by particle physics), lack of ``internal" coherence (due to the need to postulate non-baryonic Dark Matter, contrary to the overarching motivation for MONDian theories), and lack of coherence with respect to background empirical-factual knowledge (incompatibility with GWs data).

\subsection{New Relativistic MOND (RMOND)}\label{rmond}

MOND's latest relativistic incarnation is \textcite{Skordis:2020eui} ``New Relativistic Theory for Modified Newtonian Dynamics” (RMOND). \textcite[p.xxx, our emphasis]{pittphilsci19682} lauds it as ``(arguably) more successful and (arguably) less ad hoc” than the standard cosmological model: RMOND ``demonstrates that when it comes to explaining data like the CMB spectrum and the matter power spectrum, Milgromian theories […] need be no more contrived or artificial than the standard model”. Our subsequent analysis will contradict this claim.  

RMOND's starting point consists in admitting that TeVeS is observationally ruled out: ``TeVeS has been shown […] to be incompatible with the LIGO-Virgo-observations for any choice of parameters” \parencite[p.2]{Skordis:2020eui}. The data impose a ``remarkably stringent constraint" \parencite[p.1]{Skordis:2019fxt} on the arrival times of gravitational waves; gravity and light should, to very high accuracy, coincide---contrary to TeVeS (see also \textcite{Sanders:2018jiv} for an instructive analysis). 

To evade this empirical refutation, \textcite{Skordis:2019fxt} propose ``a slight generalization of TeVeS”. They extend the modification of TeVeS in \textcite{skordis2008}. In this earlier work, a family of TeVeS-like theories is constructed by cramming more terms into the action for the gravitational vector. (The action for the gravitational scalar and tensor/metric is left unchanged.) Not only does one thereby gain greater generality; one also avoids instabilities in the theory for certain applications. The original TeVeS is restored for a particular set of coefficients in those extra terms. 

In a second step, \textcite{Skordis:2019fxt, Skordis:2020eui} further modify this family of TeVeS-like theories:  they promote the coefficients of the extra terms in the action for the gravitational vector to functions of the gravitational scalar. For these coefficient functions, Skordis \& Złośnik determine a constraint---a parameter function space---such that for TeVeS-like theories respecting this constraint, ``the speed of gravity always equals the speed of light”.\footnote{\parencite[p.3]{Skordis:2020eui} subsequently also show that this holds ``even when including inhomogeneities and [that this choice] gives the same Shapiro delay as for photon".}\footnote{While this variant of the theory is constructed to pass tests related to the speed of gravitational wave propagation, it should also be mentioned that other gravitational wave tests such as those arising from observing the orbital decay of binary systems still impose significant constraints on the dynamics of the scalar sector in TeVeS-like theories \parencite{Freire:2012mg} as binary systems have long offered important constraints on many features of modified gravity theories \parencite{Will:2014kxa,Wolf:2019hun}.} Eventually, they arrive at a subset of the previous class of TeVeS-like theories by incorporating further phenomenological requirements. These ensure ``its agreement with the observed cosmic microwave background and matter power spectra on linear cosmological scales” \parencite[p.1]{Skordis:2020eui}. This class of theories they dub RMOND.   

The construction of RMOND merits a few further comments. Skordis \& Złośnik cast the previously mentioned family of TeVeS-like theories in the form of GR with two extra ingredients: (only) one gravitational metric couples to matter (exactly like in GR), with a minimally coupling Einstein-Hilbert action; but, as in TeVeS, multiple additional gravitational degrees of freedom occur, a vector and a scalar, which interact with each other and with the metric. This interaction is captured through a (complicated) free function. Its freedom is harnessed to impose constraints, explicitly designed to on the one hand, recover MONDian phenomenology, and on the other hand, to ``lead to a FLRW universe resembling $\Lambda$CDM" \parencite[p.3]{Skordis:2020eui}: the form of this function is chosen so to mimic the scaling behaviour of collisionless dust---in other words, to \textit{emulate} Dark Matter (i.e.\ the received model of Dark Matter as composed of weakly interacting, massive particles). Finally, Skordis \& Złośnik conclude their study of RMOND by investigating linear fluctuations on FLRW backgrounds; they show that RMOND indeed fits the cosmic microwave background and mass power spectra.   

Contrary to Merritt’s above-cited assessment, RMOND comes out as severely ad-hoc on the coherentist model. While at present RMOND seems empirically viable \textit{as far as cosmology is concerned} (which is a limitation in its domain, see our remarks in §\ref{laudan}), the price one has to pay for this involves multiple instances of incoherence/ad-hocness. From its origins in TeVeS, RMOND inherits TeVeS’s forms of theoretical and structural incoherence (§\ref{teves}). Here, we’ll therefore only focus on additional ones.   
Three sources of incoherence stand out.

\begin{itemize}

\item \textit{Proliferation of free parameters and functions:}

The transition from TeVeS to the \textcite{skordis2008} family of TeVeS-like theories is solely motivated by the attempt to avoid conceptual anomalies. From a MONDian perspective, no deeper justification is forthcoming. Also with respect to more general background knowledge, the transition departs from standard field-theoretical expectations. In TeVeS’s original action for the gravitational vector, ``(t)he kinetic part […] is that of an Abelian gauge  field” \parencite[p.25]{Bekenstein:2005nv}.\footnote{The gauge freedom, however, is forfeited due to the action’s other ingredients.} The inclusion of additional terms for generic TeVeS-like theories thus jettisons this familiarity. Both forms of theoretical incoherence aggravate those theories’ ad-hocness, related to the introduction of free parameters (coefficients). As a result, the TeVeS-like template for RMOND is already plagued by non-trivial ad-hocness.  

The second step of Skordis \& Złośnik’s construction of RMOND---the promotion of coefficients to functions---exacerbates this diagnosis. As discussed earlier, introducing free functions renders the ad-hocness associated with free coefficients even worse. (Note that the constraints imposed on those free functions don’t fix them; residual leeway---non-denumerably infinite degrees of freedom!---remains after imposing them.) RMOND thus displays both novel (novel, that is, vis-à-vis TeVeS!) sources of ad-hocness via proliferation of extra parameters, as well as ad-hocness via fine-tuning. 

\item 	\textit{Fine-tuning:}

Skordis \& Złośnik’s second step also lacks any inherent theoretical justification: its purpose is to tap into additional degrees of freedom to accommodate otherwise fatally conflicting empirical data.\footnote{\parencite[p.3]{Skordis:2019fxt} indeed write: ``(i)f [the coefficients] are functions of [the gravitational scalar], however, there seems enough freedom to change this fact [viz.\ conflict with gravitational wave data and Shapiro delay data].''}  Contra \textcite{pittphilsci19682} (who avers that ``the theory [RMOND] can explain those data without ‘fine-tuning’'' (2021, p.xxx)), this patently leads to fine-tuning: the constraints imposed on the free functions are designed to ensure compatibility with astrophysical and cosmological data (regarding the speed of gravitational waves, Shapiro delay, and FLRW cosmology). In this vein, one of RMOND’s inventors indeed ``admits that---unlike dark matter models that are often based on fundamental symmetry principles--the new model was not conceived with an underlying theory in mind'' \parencite{schirber_2021}. Fine-tuning----the absence of good theoretical reasons for very special features which, on the one hand, must be installed by hand, and on which, on the other hand, the theory’s success sensitively hinges---is another manifestation of ad-hocness in terms of lack of coherence (cf.\ \textcite[Ch.5]{schindler_2018}).\footnote{The introduction and significance of finely-tuned dynamical structures is a common theme in cosmological theory building and paradigm disputes. For example, similar issues also show up in the debate between inflationary vs.\ bouncing cosmologies \parencite{Wolf:2022yvd}.}

Note, by comparison, how the speed of gravitational waves follows from GR directly, without any need for additional assumptions, let alone any fine-tuning of free parameters. The same can be said about FLRW phenomenology, covered in almost every GR textbook. 
\item 
\textit{Parasitism on Dark Matter?}

We can press this line of thought further. The implementation of RMOND’s FLRW phenomenology---the requirement that it reproduce the standard (i.e. general-relativistic) cosmic dynamics---provokes a query: how convincing are the reasons, based on either general theoretical considerations or those within a MONDian framework, for imposing the constraints needed for recovering the desired FLRW phenomenology?  

One can’t help but feel that Skordis \& Złośnik’s construction is parasitic on GR. For RMOND, they prescribe an action known to reproduce ``the empirical law which concerns cosmology: the existence of sizable amounts of energy density scaling precisely as $a^{-3}$” \parencite[p.2]{Skordis:2020eui}. No reasons, general or specifically MONDian, are given for this law. Furthermore, the scalar field via which they implement this ``empirical law” is a so-called “k-essence model” \parencite{Scherrer:2004au}. This kind of model is in fact a proposal for \textit{Dark Matter}! Structurally unusual in its own right, it's regarded as a non-standard proposal for Dark Matter, and hence also itself displays lack of structural coherence with our background theoretical knowledge.\footnote{Indeed, leading cosmologist David Spergel has referred to this theory from Skordis and Z\l{}o\'snik as a ``baroque" form of dark matter \parencite{schirber_2021}.} This input seems to be ``imported” from the $\Lambda$CDM-model and its successes---a form of incoherence with \textit{MONDian} background assumptions. Consider what the leading investigators of this new version of MOND have to say on the matter. ``Within the [Dark Matter] paradigm such a law is a \textit{natural consequence} of particles obeying the collisionless Boltzmann equation. The validity of this law has been tested [...] and, during the time between radiation-matter equality and recombination, it is valid within an accuracy of $\sim 10^{-3}$” \parencite[p.2, our emphasis]{Skordis:2020eui}. Such a move rubs against the spirit of MONDian theorising as a self-standing area of gravitational research, a rival to GR in its own right.\footnote{Following \textcite{Martens2020-MARDM-13, Martens2020-MARCOT-41}, one might question the dichotomy between Dark Matter approaches and modifications of gravity (such as TeVeS or RMOND). Blurring the line between Dark Matter and modified gravity would only be grist to our mills: without a categorical distinction between those two approaches, the question becomes all the more pressing why one should prefer such a cumbersome theory like RMOND over $\Lambda$CDM if the former can only be made empirically viable by piggybacking on the latter.} If on the other hand, RMOND is classified as a(n extreme) variant of the $\Lambda$CDM-model, with exotic forms of matter (represented by RMOND’s additional gravitational variables), one might misgive: do successes on galactic scales---controversial in their significance---really justify such an inordinately unwieldy, structurally contrived theory---rather than sticking with the original $\Lambda$CDM-model? To reiterate our above caveat: little is known about RMOND outside of cosmological contexts and the MOND-regime; in particular, at present, it's not known whether RMOND can reproduce GR's successes e.g.\ with respect to, for example, gravitational wave phenomenology (besides the speed of gravitational waves) or other astrophysical applications.

\end{itemize}

\subsection{$\Lambda$CDM and Ad-Hocness}\label{lcdm}

The $\Lambda$CDM-model denotes the modern standard model of cosmology. Rather than a theory in its own right (contra Merritt’s suggestions, cf.\ \textcite[p.5]{Butterfield:2014twa}), it’s the application of GR---the standard theory of gravity---to the observed universe as a whole (under the assumption of approximate spatial isotropy and homogeneity). As its acronym suggests, characteristic of it are two assumptions:

\begin{itemize}
    \item[] ($\Lambda$) the inclusion of a non-vanishing cosmological constant, $\Lambda$, in the Einstein Equations;
    
    \item[] \textit{(CDM)} the postulate of large amounts of cold (non-relativistic) Dark Matter (abbreviated henceforth as ``CDM") with some distribution profile.
\end{itemize}

We can parse the $\Lambda$CDM-model’s potential ad-hocness into two contributions: potential ad-hocness engrained in (1) and (2), and ad-hocness engrained in the way the $\Lambda$CDM-model deals with empirical anomalies. Each will be studied separately (§\ref{4.51} and §\ref{4.5.2}, respectively). Schindler’s coherentist account yields that, notwithstanding empirical anomalies, the $\Lambda$CDM-model’s ad-hocness is moderate; it seems less severe than that of MOND or any of its relativistic extensions. 
Our subsequent analysis will not investigate GR’s standing with respect to ad-hocness. We’ll take for granted that it has ``proven its mettle”, having passed extremely diverse and precise tests (see e.g.\ \textcite{Will:2014kxa, 2014grav.book.....P}). Thus GR will be regarded as an established part of contemporary theoretical background knowledge. Likewise, we won’t examine the standing of $\Lambda$CDM-model’s underlying symmetry assumption of isotropy and homogeneity at the universe’s large-scale structures (see \textcite{Maartens:2011yx}; \textcite[p.506]{Smeenk}; \textcite{Aluri:2022re} for detailed discussions). 

\subsubsection{How Ad-hoc Are $\Lambda$CDM's Ingredients?}\label{4.51}
Within GR, the introduction of a cosmological constant is well-motivated.\footnote{Against the background theoretical knowledge of his time, even Einstein’s historical motivation was---even though it turned out to be specious (see \textcite{Earman2001-EARLTC}).}  To denounce it as an “auxiliary hypothesis” (as \textcite{pittphilsci19682, pittphilsci19913} does)--let alone a postulate ``invoked in response to \textit{falsifying} observations” \parencite[p.50, our emphasis]{Merritt:2017xeh}---would be misleading. In fact, it’s implied (and explicitly mentioned), as a \textit{free parameter of GR itself}, by both of two main heuristic approaches to GR – one found in \textcite{Weinberg1972-WEIGAC}, based on physical principles, and the other found in \textcite{Lovelock:1971yv}, based on a derivation from natural mathematical desiderata for a relativistic theory of gravity (see also \textcite[Sect.2.4.1]{Clifton:2011jh}), respectively. These compelling theoretical reasons for $\Lambda$ debunk the asserted ad-hocness of its introduction.\footnote{Of course, quantum field-theoretical contributions to the energy-matter content of the universe constitute an interesting challenge to current cosmological theorising. Including them is arguably a question of coherence with background knowledge---last not least because of the notoriously large discrepancies between predictions and observation. 
Yet, we’d like to underscore that the interpretation of $\Lambda$ as the quantum-field theoretical vacuum energy contribution shouldn’t be made automatically. As long as the discussion remains classical (as ours does), it’s perfectly consistent and plausible to treat $\Lambda$ simply as a free parameter of the Einstein equations, with no inherent connection to the quantum field-theoretical vacuum energy. 

Similar reservations apply to interpreting $\Lambda$ as the present-day manifestations of a dynamical scalar, i.e. an extra field (see e.g.\ \textcite{Li:2011sd} for a review). Such a so-called “Dark Energy” interpretation shouldn’t be made automatically either, as e.g.\ \textcite[p.118]{Merritt2020} does. \textcite[p.25]{pittphilsci18810} rightly observes: ``Merritt confuses the cosmological constant with dark energy”. The important point is that the evidence for the cosmological constant is very strong, while there is no evidence for any more complicated form of dark energy”.}

They are complemented by empirical reasons. High-precision data mandate a non-vanishing value for $\Lambda$. The $\Lambda$CDM-model naturally and satisfactorily accommodates\footnote{Chan rightly stresses that the $\Lambda$CDM-model’s ability to accommodate the data isn’t guaranteed a priori; that GR, with a non-vanishing cosmological constant, achieves the best-fit for current cosmological data is an impressive achievement \parencite[p. 286]{Chan2019-CHAACO-33}.} variegated types of observations (see e.g.\ \textcite[p.734]{pittphilsci16893}; \textcite{Ishak:2018his} for a comprehensive technical review), such as: 

\begin{itemize}
    \item The temperature anisotropies in the CMB data allow cosmologists to determine both the universe’s geometric shape, as well as the total density of matter. From both components, the contributions of a cosmological term $\Lambda$ can be computed.  
	\item The light of far-distant supernovae attenuates more rapidly than one would anticipate. This observation is most naturally interpreted as evidence of the universe’s accelerated expansion. The $\Lambda$-term acts repulsively, like a fluid with a negative pressure, propelling the expansion.
	\item The length-scale given by the angular positions of the acoustic peaks in the CMB anisotropy map (the so-called ``acoustic baryonic oscillations”) have left an imprint in the distribution of large-scale structures. By comparison with the distribution of galaxies, cosmologists can determine the expansion history of the universe. Again, the results are most naturally construed in terms of an accelerated universe, corresponding to a non-vanishing cosmological constant.
\end{itemize}

 $\Lambda$’s occurrence per se---no matter how well-motivated---is a source of ad-hocness (a form of GR’s inherent ad-hocness, that is). The smallness of $\Lambda$‘s numerical value (as determined from the data), in particular, has disconcerted some (see \textcite{Nobbenhuis:2006yf} for a review) as ``unnatural” (for a, to our minds, convincing counter, see e.g.\ \textcite{Hossenfelder2019-HOSSFE}). Fortunately, we may steer clear of that debate: with respect to both the empirical need for $\Lambda$ as well as its peculiar numerical value, the $\Lambda$CDM-model isn’t worse off than MONDian theories; they too must account for the data---by incorporating a $\Lambda$-term or something similar (in TeVeS via a suitable choice for its free function). 

In contradistinction to GR, however, the above-cited natural motivation for this addition lapses for MONDian theories: to the best of our knowledge, relativistic extensions of MOND can’t be derived from physical or mathematical principles (cf.\ \textcite[p.1]{Skordis:2020eui}). The verdict regarding $\Lambda$’s ad-hocness is thus reversed: whereas the $\Lambda$CDM-model \textit{naturally} (viz.\ through a free parameter of the theory itself) accommodates a diverse and precise data set, MONDian theories require an ad-hoc adjustment, the \textit{theoretically unmotivated} stipulation of a cosmological constant. 

What about the standard cosmological model’s CDM\footnote{We’ll not discuss the assumption that Dark Matter has to be ``cold” (i.e. moves at non-relativistic speeds). The presently predominant preference for cold dark matter has to do with structure formation \parencite{Dodelson:1995es}: ``hot” dark matter (light particles, moving at relativistic speeds, such as neutrinos) would wash out the seeds for galaxy formation.} component? How ad-hoc, on the coherentist model, is the postulate of Dark Matter as a form of non-luminous matter? We begin with rebutting the claim (championed e.g.\ by \textcite{Kroupa:2012qj} or \textcite{Merritt:2017xeh}) that the inclusion of CDM is a (methodologically reprehensible) dodge to evade falsification. Read at face value, the claim is plainly false: no physical principle dictates that, on pain of falsification, all matter must be visible/luminous. Pointing to the history of speculations about invisible matter (cf.\ also \textcite{Bertone:2018krk}), \textcite[p.286]{Chan2019-CHAACO-33} reminds us: “(b)y definition, dark matter is a matter that cannot be observed using electromagnetic radiation. […] Generally, speaking, cosmologists use the term ‘dark matter’ to represent those dark bodies and unknown matter in the universe […]. […] We have acknowledged the existence of active neutrinos […] since 1956, which are a component of dark matter […]”. Neutrinos simply \textit{are} an empirically verified form of Dark Matter in the sense of a non-luminous type of matter.   

Perhaps then, the claim targets something more specific: the belief that ``matter detected only via gravitational effects is somehow an addition to GR\footnote{See e.g.\ the explicit charge in \textcite[p.113]{sanders_2016}.} […]. However, GR says nothing about the sources of the gravitational field. […] Thus, the claim that some new sort of matter, no matter how it is inferred, somehow falsifies the standard model is certainly untrue if referring to GR” \parencite[p.11]{pittphilsci18810}. GR asserts a law-like relation between gravity and the energy-stress, generated by matter. A matter theory (e.g.\ electromagnetism or a mechanical model of dust) supplies crucial information about the latter.\footnote{Einstein stressed this feature from early on \textcite{LEHMKUHL2019176}, emphasising in particular the phenomenological nature of the matter-theoretic input (in both his and our times).} Assumptions about Dark Matter are just a particular kind of such input. Without matter-theoretic input, GR---a theory of \textit{gravity}, not of its sources (i.e. matter)---is incomplete. But such incompleteness or non-comprehensiveness, the need for a complementing matter theory, clearly differs from the charge of ad-hocness. But would that count as a bug of GR? Or should instead we regard it as a feature we’ve got to live with? The answer arguably depends on (ipso facto) controversial, antecedent metaphysical commitments about Nature’s unity: does reality admit of one fundamental comprehensive theory (see e.g.\ \textcite{Cat:1998rka, Hossenfelder:2018jew} for critical voices from the physics community)? We’ll not embroil ourselves in this conundrum (last, but not least because it seems to equally apply to MONDian theories) and move on.

Does some form of ad-hocness then lurk instead in the \textit{manner in which such matter input is postulated}? Indeed, while hitherto attempts to experimentally detect Dark Matter candidates have failed (e.g.\ \textcite[Sect.IX]{Bertone:2018krk}), ``usually the existence of what has been at least initially been perceived to be dark matter was later confirmed by non-gravitational means” \parencite[p.11]{pittphilsci18810}. Before tackling the issue of empirical evidence---and hence potential ad-hocness in terms of deficient coherence with background empirical knowledge---let’s inspect the method of inference of, our epistemic access to, Dark Matter: is there anything methodologically problematic about the fact that Dark Matter, at present, is inferred only through its gravitational effects? 

The history of astronomy attests to the wide-spread use of this inferential practice (see e.g.\ \textcite{SmithG}). Also today, ``(m)ost cosmologists treat these developments [inferring Dark Matter via gravitational effects] as akin to Le Verrier’s discovery of Neptune. In both cases, unexpected results regarding the distribution of matter are inferred from observational discrepancies, using the theory of gravity” \parencite[p.518]{Smeenk}. Reassuringly, the charge can be dispelled: what matters for methodologically sound acceptance of hypotheses is independent testability. That requires that the gravitational effects of Dark Matter play a role in different contexts (e.g.\ gravitational lensing \text{and} effects on visible objects in the vicinity). As long as we can experimentally probe those contexts, independent testability is vouchsafed (op.cit., p.509).\footnote{Other means of access to Dark Matter effects, via (hypothetical) different types of interaction would arguably only make it easier to independently test hypotheses about Dark Matter---a question of practicality, not of principle. Absence of such other interactions doesn’t per se preclude independent tests.} 

Indeed, this seems the case for Dark Matter. Evidence for it has accumulated from a plethora\footnote{NB: In the main, this multiplicity of evidence for Dark Matter only arises, so long as one doesn’t neglect the high-precision supra-galactic and cosmological data. This data is indeed largely accepted in the scientific community; hence the overwhelming acceptance of the existence of Dark Matter. Conversely, it's not surprising that \textit{by neglecting} it, \textcite{Vanderburgh:2003pvz}'s scepticism towards Dark Matter appears much more alluring than most physicists would grant. (Ironically, \textcite{Vanderburgh2014-VANPAN-2} also underestimates the historical importance of cosmological evidence for the reception of the Dark Matter problem.)} of high-precision data (see e.g.\ \textcite[p.732]{pittphilsci16893}; \textcite[Ch.1]{Anderl}), such as:

\begin{itemize}
    \item As repeatedly referred to earlier, rotation curves of single disc galaxies flatten out in a way that contravenes what one would expect given the observable/luminous matter distribution, and Newton’s law of gravity. The “mass discrepancy” is resolved by postulating missing Dark Matter.
    \item Larger structures, such as galaxy cluster, exhibit a similar “mass discrepancy”: in order to account for the structure’s observable dynamics (i.e. the behaviour of visible masses), one has to postulate missing Dark Matter.
    \item Gravitational lensing (especially in colliding galaxy clusters) indicate large amounts of non-luminous matter: they are responsible for the observed, otherwise enigmatic bending of light from background stars.
    \item The temperature anisotropies in the CMB are too small to facilitate the formation of larger structures, such as galaxies. Dark Matter comes to rescue – as a non-baryonic addition to ordinary/baryonic matter: its gravitational pull allows baryonic matter to clump sufficiently. \item As before in the case of $\Lambda$, the CMB contains information about the universe’s total energy content and density (via the position and height of the first peak in the CMB’s power spectrum, respectively)---as well as information about the contributions of ordinary/baryonic matter (via the height of the second peak). Again a “mass discrepancy” gapes, resolved by postulating missing mass in the form of non-baryonic Dark Matter. 
\end{itemize}

\textcite[p.511]{Smeenk} sums up the evidential situation: “(t)he consistent determination of these parameters [characterising the $\Lambda$CDM-model] from many different types of observations supports an overdetermination argument much like Perrin’s. […] the $\Lambda$CDM model leads to systematic connections between a diverse array of observable features of the universe. \textcite{Peebles:2004qg}, for example, enumerates 13 distinct ways of measuring the overall matter density $\Omega_0$ at large scales: several distinct techniques based on using galaxies as mass tracers; weak lensing; cluster mass functions; the mass fluctuation power function; and so on.” 

A ``remarkably consistent picture of the world as probed by several independent tests” emerges \parencite[p.63]{sanders_2016}. Even the self-styled “dedicated sceptic”---and advocate of MOND---Sanders “(has) to admit that these independent lines of evidence […] provide strong evidence for the presence of dark energy and dark matter that produces predictable effects on a cosmological scale” (ibid, p.64). Successfully accommodating and unifying a multitude of phenomena, the $\Lambda$CDM-model achieves non-trivial coherence with background empirical-factual knowledge.\footnote{This isn’t to deny, as \textcite[p.64]{sanders_2016} rightly stresses, “tensions in the cosmological data. The significant difference between the Hubble parameter as measured from the CMB anisotropies and that implied by the supernovae observations is a clear indication of systematic effects in one or both data sets”. Another problem is the lithium problem, mentioned earlier (see e.g.\ \textcite{PhysRevLett.117.152701}, cf.\ \textcite[Ch.6]{Merritt2020} for a useful (albeit biased) exposition of the problem, and \textcite[Sect.4.1]{Chan2019-CHAACO-33}, and \textcite[p.95]{Anderl} for a measured philosophical response).}

Yet, some issues remain, empirical as well as theoretical-conceptual ones. They are, as we argued, best construed as open questions or anomalies---challenges but not counterevidence. Let’s finally turn now to theoretical-conceptual issues: how does the postulate of Dark Matter cohere with background \textit{theoretical} knowledge?

Here, the $\Lambda$CDM-model is dealt a hard blow: the standard model of particle physics, the default matter theory, \textit{doesn’t} contain any suitable candidate for Dark Matter (of the type that the above-cited evidence points to). To-date, the hunt for independent experimental confirmation of Dark Matter has been futile. Established physics doesn’t cohere with the $\Lambda$CDM-model. In this regard the Dark Matter hypothesis comes out as indeed ad-hoc. 

The following considerations, however, mitigate that verdict. For this, we should differentiate between the evidence for Dark Matter and the concomitant inference to some kind of gravitating matter on the one hand, and specific Dark Matter candidates (or models) on the other.\footnote{Importantly in this regard, \textcite[p.15]{Bertone:2016nfn} clarify: ``this phrase [Dark Matter] is most frequently used as the name, a proper noun, of whatever particle species accounts for the bulk of our Universe’s matter density.”}  As discussed above, the former can be acquitted of the charge of ad-hocness: inferring Dark Matter from the observational evidence not only harmonises with, but is impelled by, our standard background theoretical knowledge, namely GR (cf.\ \textcite{pittphilsci16893} for a complementary analysis). It simply follows from the best theory of gravity available---even at the (manageable) cost of some ad-hocness in the form of postulating otherwise poorly understood, novel matter.\footnote{\textcite[Ch.1 and Ch.9.1]{sanders_2016} reproaches the ``essentially reductionist predilection in contemporary science” (p.9), which demands that ``all follows from a few basic laws or a ``’theory of everything’ and to derive such a theory has become a modern holy grail” (ibid.). We are sympathetic to Sanders’ scepticism about ``theories of everything". Nonetheless, Sanders’ criticism strikes us as a red herring. In the case of the $\Lambda$CDM-model the issue is whether we are justified in relying on our best theory of gravity, corroborated in multiple, strict test---rather than some MONDian alternative. As far as ad-hocness in terms of coherence with established theoretical background knowledge is concerned, the answer is to favour the $\Lambda$CDM-model.}  With respect to ad-hocness, the status of specific Dark Matter candidates, by contrast, is more delicate. 

They are propounded to address ``the principal problem of the paradigm [i.e. the $\Lambda$CDM-model]: the mysterious nature of the two dominant components---dark energy and dark matter” \parencite[p.4,2,47,64]{sanders_2016}. Via such models, one seeks to integrate Dark Matter into our theoretical background knowledge---by expanding this background knowledge, thanks to (hopefully) warranted extensions of the standard model of particle physics. The problem here is ``not a lack of viable suggestions, i.e. models, concerning the nature of dark matter, but their abundance” \textcite[p.3]{Martens2021-MARDMR-5}. Rather, the problem consists in lack of independent empirical support for any of them: empirical constraints to-date fail to reduce ``this myriad of possibilities” (ibid.).\footnote{In light of this severe underdetermination by evidence, \textcite{Martens2021-MARDMR-5} emphatically counsels epistemological caution when it comes to committing to any \textit{specific} proposal for Dark Matter in particular---a view with which we agree. } This underdetermination is only transient, though; it’s reasonable to assume that future data will overcome it. In this sense, the promise of Dark Matter models hasn’t been redeemed (yet). 

At the same time, the non-trivial theoretical/super-empirical credentials of many of those Dark Matter models must be underlined. Two of the most popular Dark Matter candidates are cases in point: particles postulated by so-called super-symmetric extensions of the standard model, and axions, respectively. Both ideas were originally devised independently of the Dark Matter context: they were intended as resolutions of shortcomings (or blemishes), afflicting the standard model of particle physics. Independent theoretical reasons for these proposals thus allay the charge of ad-hocness (cf.\ see \textcite[Ch.9]{Profumo} for a detailed discussion of theoretical (non-)ad-hocness of Dark Matter candidates)---notwithstanding the absence of empirical corroboration. That is, thanks to their independent motivation, those Dark Matter proposals cohere with background beliefs---\textit{even if}, at present, these beliefs remain empirically unwarranted (or, as in the case of the most natural super-symmetric Dark Matter candidate, have even come under pressure from persisting null results).\footnote{Herein---in terms of coherence with background beliefs---lies the critical difference between the lack of experimental warrant for Dark Matter proposals, and the absence of a relativistic version of MOND: for better or worse, the ideas motivating those proposals---\textgreek{ὡς ἐπὶ τὸ πολύ}---belong to the scientific community's background beliefs. By contradistinction, relativistic extensions of MOND either squarely conflict with our background beliefs about facts (as in the case of TeVeS), or we lack independent justification for their construction (as in the case of RMOND)---or, as in the case of hitherto uninvented MONDian theories, for their very existence. We thank an anonymous referee for pressing us on this!}   

We conclude that with respect to coherence with theoretical background knowledge the $\Lambda$CDM-model’s score is mixed: the postulate of Dark Matter itself isn’t ad-hoc; by contradistinction, attempts to reconcile it with our matter-theoretical background knowledge through specific Dark Matter models are---yet, in virtue of their non-trivial theoretical/extra-empirical motivation, perhaps not egregiously so.

\subsubsection{Succumbing to Ad-hocness in the Face of Anomalies?}\label{4.5.2}

As alluded to in the foregoing, the $\Lambda$CDM-model isn’t free of empirical challenges. Like all theories, it’s engulfed in an ``ocean of anomalies” or ``recalcitrant instances”. Do they tarnish the $\Lambda$CDM-model with ad-hocness, as e.g.\ \textcite{Merritt:2017xeh} avers? 

According to \textcite[p.125]{sanders_2016}, ``the essential problem with the paradigm [the $\Lambda$CDM-model] is that cosmology, via dark matter, impinges directly upon the dynamics of well-studied local systems---galaxies---and here […] the cosmological paradigm fails.” Dark Matter should cluster locally on galactic scales. ``[…] $\Lambda$CDM fails on this scale […]”. These alleged failures are precisely the phenomena where MOND scores successes (ibid., Ch.8). 

As pointed out earlier, Sanders’ assessment doesn’t jibe with the present-day consensus. According to the latter, it would be exaggerated, or at least premature, to brand these phenomena as empirical refutations. Rather than inconsistencies between the data and the $\Lambda$CDM-model, they only demonstrate “failures” of extant astrophysical simulations. Of necessity, these involve very complex astrophysical and nuclear-physical processes; many  details and mechanisms here remain ill-understood. Hence, those phenomena should best be viewed as unresolved challenges or yet undigested anomalies---a Kuhnian “puzzle” (see also \textcite[Ch.3.2]{Anderl}; \textcite[Sect.5.2]{DeBaerdemaeker2022-DEBMAM-2}). Accordingly, judgement as to their (counter-)evidential ramifications for the $\Lambda$CDM-model ought to be suspended.\footnote{Conversely, \textcite[p.9]{pittphilsci18810} rightly warns against false triumphalism: “proponents of the $\Lambda$CDM-model shouldn’t simply claim, without evidence, that more complicated simulations will explain all observations.” 
In the same vein, he “(thinks) it is fair to say that the primary astrophysical motivation for dark matter is ad hoc” (ibid.). Insofar as Helbig wants to stress (1) that the rotation curves constitute only one argument in favour of Dark Matter, and (2) that on its own---that is, if it were the only argument---it wouldn’t be particularly compelling (especially in light of a predictively successful rival, viz.\ MOND), we concur. Needless to say, (2) is counterfactual: astrophysical evidence for Dark Matter at galactic scales is complemented by cosmological evidence. The persuasiveness of the $\Lambda$CDM model’s Dark Matter postulate derives from the combination, or confluence, of evidence. 
This matches how historically the consensus regarding Dark Matter formed: it was the realisation that the cosmological desire (with primarily philosophical and super-empirical support at the time) and the empirical astrophysical anomalies admit of a single solution---the postulate of Dark Matter (\textcite{deSwart:2017heh, deSwart:2019bsf}).} \textcite[p.265]{Peebles2020} captures this attitude (cf.\ \textcite{DeBaerdemaeker2020-DEBJSS}, whose philosophical analysis fleshes out a similar conclusion, and \textcite{Scott:2018adl}, who offers another view from the physics community in alignment with Peebles): ``(t)he community assessment is that this [i.e. the successes of MOND] is an accident of the complexity of the application of standard physics to galaxy formation. Deciding whether we have adequate physics for analyses of the structures of galaxies, […], or whether we have missed something interesting, calls for more data analyzed in better ways, as usual. Meanwhile the community decision is appropriate: work with standard physics and the hypothetical subluminal/nonbaryonic matter applied to a cosmology that fits demanding tests—until or unless we run into trouble.” In conclusion, we cannot discern any particularly worrisome form of ad-hocness in the way the $\Lambda$CDM model handles anomalies. 

\section{Summary}\label{summary}
\noindent Let's summarise the main findings that we sought to establish in this paper. Section §\ref{method} furnished a corrective to claims in the MOND-related literature that MONDian research receives a particularly favourable appraisal within the classical methodological frameworks. More specifically:
\begin{enumerate}
    \item From a Popperian perspective, MONDian theories have either been falsified or are methodologically inferior. Nonetheless, MOND’s predictive successes in a domain fraught with on-going debate create an exciting problem-situation. 
\item MOND doesn’t count as a Kuhnian paradigm. Kuhn’s analysis of theory virtues as guides to theory choice elucidates why the physics mainstream hasn’t adopted MOND. 
\item MONDian research doesn’t constitute a research programme in Lakatos’ sense. In particular, it lacks a positive heuristic. The progressive problem-shift within MONDian research was found to be overstated. 
\item MONDian research falls within \textit{GR’s} research tradition (rather than a research tradition in its own right), as envisaged by Laudan. This explains MOND’s marginality in the physics community---both with respect to its acceptance and its pursuit.  
\end{enumerate}

Section §\ref{adhoc} explicated the main charge against MOND, its ad-hocness, via Schindler’s coherentist model. MOND was found to achieve substantial coherence with respect to background factual knowledge on galactic scales; it conflicts with theoretical background knowledge. MOND’s only empirically viable relativistic version, whilst inheriting MOND’s improvements with respect to background factual knowledge on galactic scales, displays significant lack of coherence, both internal/structural as well as with respect to theoretical background knowledge; its coherence---or even consistency---with established factual knowledge outside of the MOND regime and the CMB data is presently unknown. A comparison with the $\Lambda$CDM model’s ad-hocness turned out in favour of the $\Lambda$CDM model. An important exception concerns its lack of coherence with respect to theoretical background knowledge regarding the nature of Dark Matter. This result is unsurprising: one expects Dark Matter to burst the boundaries of established knowledge—and accordingly conservativism, on which aversion to ad-hocness pivots, will be stretched to its limits. This, after all, makes Dark Matter such a---philosophically and physically---thrilling topic!

\section*{Acknowledgements}

\noindent Many thanks for feedback on previous versions of the manuscript from our friends and colleagues: Brian Pitts, Yemima Ben-Menahem, Samuel Schindler, Raz Chen-Morris, and John Norton. The authors acknowledge financial support through PD's "Eigene Projektstelle des Landes Bremen" (AG Sieroka). PD is grateful for support through the Martin Buber Society of Fellows for Research in the Humanities and Social Sciences, Hebrew University of Jerusalem, IL. 

\printbibliography

\end{document}